\documentclass[11pt,english,sort&compress]{article}
\usepackage[T1]{fontenc}
\usepackage[latin9]{inputenc}
\usepackage{geometry}
\usepackage{textcomp}
\usepackage{amsmath}
\usepackage{amssymb}
\usepackage{graphicx}
\usepackage{esint}
\PassOptionsToPackage{normalem}{ulem}
\usepackage{ulem}

\makeatletter
\newcommand{\lyxaddress}[1]{
\par {\raggedright #1
\vspace{1.4em}
\noindent\par}
}

\makeatother

\usepackage{babel}
\begin{document}

\title{Topical Review\\
 Spatio-temporal phenomena in complex systems with time delays}

\author{Serhiy Yanchuk$^{1}$ and Giovanni Giacomelli$^{2}$}

\maketitle

\lyxaddress{$^{1}$Technische Universität Berlin, Institute of Mathematics, Straße
des 17. Juni 136, 10623 Berlin, Germany }

\lyxaddress{$^{2}$CNR-Istituto dei Sistemi Complessi, via Madonna del Piano
10, I-50019 Sesto Fiorentino (FI), Italy}

\maketitle 

\tableofcontents{}
\begin{abstract}
Real-world systems can be strongly influenced by time delays occurring
in self-coupling interactions, due to the unavoidable finite signal
propagation velocities. When the delays become significantly long,
complicated high-dimensional phenomena appear and a simple extension
of the methods employed in low-dimensional dynamical systems is not
feasible. We review the general theory developed in this case, describing
the main destabilization mechanisms, the use of visualization tools
and commenting the most important and effective dynamical indicators
and their properties in different regimes. We show how a suitable
approach, based on the comparison with spatio-temporal system represents
a powerful instrument to disclose the very basic mechanism of long
delay systems. Various examples from different models and a series
of recent experiments are reported. 
\end{abstract}

\section{Introduction}

The utmost important missions of nonlinear dynamics are certainly
the understanding of the working principles of complex real-world
or artificial systems, and predicting and controlling their behaviors.
Several dynamical systems have proven their surprising effectiveness
in modeling specific experiments and natural phenomena; the most common
belong to the class of ordinary differential systems (ODS), discrete-time
systems, or partial differential equations. While the first two are
finite-dimensional and determine the dynamics on the basis of finitely
many variables, the latter is infinite-dimensional, and it is commonly
employed to describe the evolution of spatially-extended systems

Another extremely useful mathematical tool is represented by delay-differential
systems (DDS), which play a special role within the toolbox of dynamical
systems. Starting from the sixties of the last century up to now,
the use of such systems spread into practically all possible application
fields, ranging from biology and neuroscience \cite{Mackey1977a,Popovych2006,Izhikevich2006},
laser physics \cite{Ikeda1979,Erneux2009,Soriano2013,SciamannaShore2015},
to population dynamics\cite{Kuang1993}, quantum physics \cite{KopylovEmarySchoellEtAl2015},
and traffic systems \cite{Orosz2010} just to name a few. In contrast
to finite-dimensional systems, DDS take into account the time lags,
which appear naturally in realistic models when properly keeping in
account the finite propagation times of the signals, finite reaction
times or velocity of bodies. An exemplary situation is represented
by the retarded potentials in classical electrodynamics \cite{Page1918,Jackson1975};
their effect can also lead to some surprising consequences \cite{DeLuca1998}.

In 1977, Mackey and Glass introduced a simple-looking, scalar physiological
control system with time delay showing surprisingly rich dynamical
behaviors including deterministic chaos \cite{Mackey1977a}. Two years
later Ikeda proposed a delay model for the ring laser \cite{Ikeda1979},
which was very popular till nowadays. Apart from their elegance and
apparent simplicity, the success and importance of these models was
incepted by a proper timing. At that time, research on deterministic
chaos was rising, and the above mentioned models were among the utmost
examples of simple-at-a-glance, time-continuous scalar systems exhibiting
chaotic dynamics. A further, very important model was suggested by
Lang and Kobayashi \cite{Lang1980} in 1980 for a semiconductor laser
with delayed feedback, and it appeared to be extremely useful in describing
the dynamics even on a quantitative level in some situations.

In all the aforementioned systems, the time delay is a parameter of
paramount importance. The early study of the qualitative properties
of chaotic attractors in DDS, and especially their dependence on time
delay, was carried out by Farmer in \cite{Farmer1982}. The mathematical
analysis of DDS can be dated back at least to 1949 \cite{Myshkis1949},
since then, it has been developed significantly \cite{Bellman1963,Kuznetsov1982,Hale1993,Diekmann1995,Atay2010,GuoWu2013}.
Early mathematical results on systems with possibly large time delays,
relevant for the purpose of this review include the contributions
of Mallet-Paret and Nussbaum on systems with monotone delayed feedback
\cite{Mallet-Paret1976,Mallet-Paret1986}, in particular, the appearance
of square waves. Kashchenko et al. \cite{Grigorieva1992,Grigorieva1993a,Grigorieva1993b,Kashchenko1998,Kashchenko2016}
reported about the formal asymptotic reductions of multiscale delay
systems to partial differential equations.

The simplest DDS with a single time delay $\tau$ has the form 
\begin{equation}
\frac{dx}{dt}(t)=F(x(t),x(t-\tau)),\label{general}
\end{equation}
similar to a system of ordinary differential equations\footnote{Here we do not consider the situation when the derivative in (\ref{general})
is delayed. Such systems are called neutral and possess several properties
that differ them from (\ref{general}), see \cite{Hale1993}.}; however, the presence of the time delay implies that its phase space
is infinite-dimensional. Indeed, this fact was pointed out already
by the very first researches \cite{Vogel1965}. In particular, an
initial condition for this system should be given as a function on
the interval of length $\tau$, i.e. $x_{0}(\theta)$, $\theta\in[-\tau,0]$,
if evolution $x(t)$ is to be found for $t>0$. Such segments of function
of length $\tau$ can be considered as points in the infinite-dimensional
phase space, and the system (\ref{general}) describes the evolution
in such a phase space. Accordingly, the notation $x_{t}$ is sometimes
used in the mathematical literature to denote such a ''point'' at
time $t$: $x_{t}:=x_{t}(\theta),\,\theta\in[-\tau,0]$. 

DDS are infinite-dimensional even for very small delay. This statement
would seemingly lead us to a contradiction with a common sense that
very small time delays should not make any effect on the dynamics,
and, thus, the system in such a case should be finite dimensional.
This contradiction is resolved by the fact that DDS, being formally
infinite dimensional systems, exhibit a dynamics which is essentially
\emph{finitely} dimensional. Using a more precise mathematical language,
this means that the system's dynamics $x(t)$ approaches a certain
finite-dimensional attractor or manifold. For small time delays $\tau\to0$,
this can also be shown rigorously \cite{Chicone2003}. In such a case,
the dynamics on the manifold is governed by the reduced finite-dimensional
system $x'=F(x,x)$. As the time delay increases, the complexity often
increases as well as the dimensionality of the dynamics \cite{Farmer1982,Kotomtseva1985,LeBerre1986,Ikeda1987,Dorizzi1987,LeBerre1987,Giacomelli1989,Lepri1993,Giacomelli1995,Lee2004,ChemboKouomou2005,Lavrov2009}
leading to new phenomena and increased coexistence of different attractors.
Meanwhile, it is commonly accepted that the effective dimensionality
of the systems can become arbitrary large as the delay grows \cite{Arecchi1992,Lepri1993,Giacomelli1994,Giacomelli1998,Wolfrum2006,Yanchuk2009,Yanchuk2010a,Yanchuk2015a}. 

The following points stay at the very basis of the message of this
review:
\begin{enumerate}
\item Time delay systems can exhibit dynamical phenomena that can be arbitrary
high-dimensional. A more precise meaning of this will be made clear
in the subsequent sections of the article. 
\item The dimensionality of the observed phenomena is proportional to the
time delay $\tau$. 
\item A proper description of the high-dimensional dynamics in DDSs requires
methods, that are different from those used for low-dimensional ODE
systems. In particular, the normal forms that describe the destabilization
processes are sometimes similar to those for spatially extended systems
(PDEs). 
\end{enumerate}
Along the review, we will try to clarify the above general statements
and make them more precise, as well as to present a variety of realizations
of the spectacular high-dimensional phenomena exhibited by DDS, their
theoretical analysis and experimental applications. 

In order to make this paper more focused, we decided not to treat
the role of noise on the dynamics. It is widely considered in the
literature (see e.g. the special issue \cite{SpecialIssue2005}) and
it would deserve a separate detailed discussion.

The structure of this manuscript is the following. In Sec.~\ref{sec:long delay systems and st dyn},
we introduce the spatio-temporal representation, a useful method to
visualize the solutions $x(t)$ of systems with large delays or data
from feedback experiment with a large time lag. We also explain the
physical meaning of this tool using the autocorrelation function.
Subsection.~\ref{sub:maps} reviews shortly the literature on discrete
maps with large delays. Finally, we mention a formal, but significant
interpretation of nonlinear delay systems as composition of two operators:
the nonlinear function and a linear differential evolution scheme. 

Section \ref{sec:LyapunovSpectrum} describes the properties of various
complexity measures and their scaling with time delay: Lyapunov spectrum,
Floquet exponents, Kaplan-Yorke dimension, entropy, as well as the
number of coexisting periodic solutions. Taking into account that
the spectrum converges asymptotically to a continuous shape, the classification
of instabilities is presented in Sec.~\ref{sub:Classification-of-instabilities}
for large-delay systems, showing the similarity with that for spatially
extended systems.

In Sec.~\ref{sec:math}, the applications of the multiple scale analysis
to large-delay systems are given. In particular, it is explained how
the normal forms can be derived for each particular type of instability.
The physical discussion about the long-delay limit is provided in
Sec.~\ref{sec:phys-mean}, where information propagation aspects,
drift in the spatio-temporal representation, and boundary conditions
are discussed. 

Section~\ref{sec:examples-theory} is devoted to the several spatio-temporal
phenomena reported in the context of high-dimensional delay systems,
such as spatio-temporal chaos, Eckhaus instability, stripes and square
waves, strong and weak chaos, coarsening, nucleation, localized structures,
spiral defects and turbulence. 

The next Sec.~\ref{sec:examples-exp} reports on the experiments
on long-delayed systems, which are known to be mostly optical systems
till now. Finally, we conclude with a summary and a discussion of
future perspectives in Sec.~\ref{sec:concl}.

\section{Long-delay systems and spatio-temporal dynamics\label{sec:long delay systems and st dyn}}

The DDS (\ref{general}) is an infinite-dimensional system, where
the state is determined by the functions $x_{t}(\theta)$ , $-\tau\le\theta\le0$
that can be considered as ''points'' in the phase space. It is therefore
natural to interpret the time interval of length $\tau$ as the \textit{space(-}like)
domain and try to map the problem to an evolution rule for the functions
defined on this domain. A straightforward way of introducing the spatio-temporal
coordinates is as $u(\theta,t)=x_{t}(\theta)=x(t+\theta)$, where
$\theta$ plays the role of a spatial coordinate and $t$ of a temporal
one. In this case, the resulting dynamics is described by an equivalent
transport equation (hyperbolic PDE with complicated boundary conditions),
see Sec.~\ref{sub:hyperbolicPDE} and system (\ref{eq:hyperbpde}).
However, the spatial variable defined in this way is, in fact, just
a re-parametrization of the time $t$ and the obtained PDE does not
provide an additional physical insights into the dynamics. Therefore,
we do not consider this representation except for a short description
in Sec.~\ref{sub:hyperbolicPDE}. 

Instead, here we concentrate on another spatio-temporal representation
of the dynamics of system (\ref{general}), when $\tau$ becomes large
and the temporal and spatial variables can be defined in such a way,
that they decouple from each other to some extent. The reason behind
such a decoupling is a difference in time scales, on which the new
spatial and temporal variables are acting. We hope that this statement
becomes more clear after the reading of the following sections. As
a result of this representation, the obtained spatio-temporal system
is simpler (usually of parabolic type) and allows to obtain additional
insights in the properties of the system. Moreover, different systems
with feedbacks, when the propagation times in the feedback loops are
significantly larger than their natural time-scales, show common features
that can be described in a unified way leading to some kind of normal
forms. 

In this section, we introduce and discuss the main ideas behind the
spatio-temporal representation, focusing on its physical interpretation;
more general comments on the consequences of the method are presented
in Sec.\ref{sec:phys-mean} while n Sections \ref{sec:LyapunovSpectrum}
and \ref{sec:math}, more theoretical details and derivations will
be given.

\subsection{The spatio-temporal representation\label{sub:str}}

The heuristic equivalence of DDS's with 1D, spatially extended systems
can be pushed further with the aid of the following example, based
on the Stuart-Landau delayed equation \cite{Giacomelli1994}:

\begin{equation}
\frac{dx}{dt}(t)=\mu x(t)+\nu|x(t)|^{2}x(t)+\eta x(t-\tau).\label{eq:sl}
\end{equation}
System~(\ref{eq:sl}) was introduced to describe phenomenologically
the oscillations of the emitted intensity from a $CO_{2}$ laser,
in a setup with delayed, optoelectronic feedback on the losses. It
is a modification of the normal form equation for the Andronov-Hopf
bifurcation, with an extra additive, linear delayed coupling to keep
in account the feedback at the lowest order. 

\begin{figure}
\centering{}\includegraphics[width=1\linewidth]{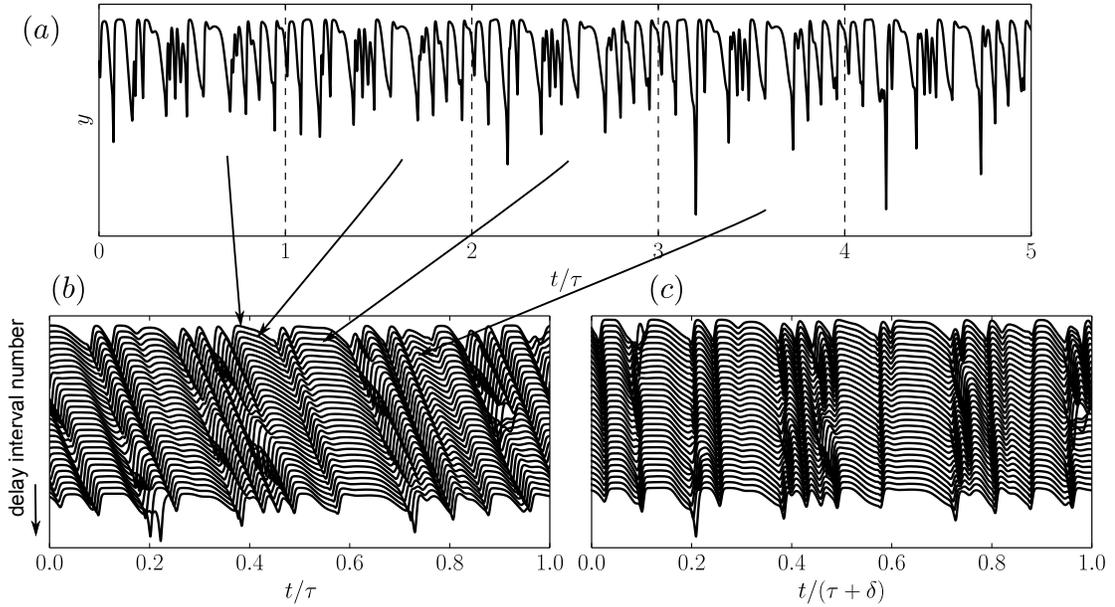} \caption{\label{fig:STR-example}Illustration of the spatio-temporal representation
and the drift. A trajectory of system (\ref{eq:sl}) with $\tau=100$,
$\mu=1+i$, $\eta=1.2$, and $\nu=-0.1-0.22i$ is shown in panel (a).
Panel (b) shows consecutive pieces of length $\tau$ of the trajectory
plotted one under another. (c) shows the same figure as (b) but with
a slightly enlarged lengths $(\tau+\delta)$ of the intervals. For
an appropriately chosen $\delta$ (the drift), the vertical correlation
is maximal.}
\end{figure}

Let us consider the temporal series depicted in Fig.\ref{fig:STR-example}(a),
obtained from a numerical integration in a chaotic regime with a delay
time $\tau\gg1$. Partitioning the time series in slices of length
$\tau$ (with an arbitrary choice of the initial point), it appears
that a certain degree of coherence is maintained between the value
of the solutions at corresponding points (times) within consecutive
slices. To better evidence such a behaviour, the slices can be plotted
in a two-dimensional arrangement (Fig.\ref{fig:STR-example}(b)).
The horizontal axis can now be naturally interpreted as a \textit{space}-like
variable, indexing the variable spanning a delay-long cell. The role
of a \textit{time}-like variable is played by the integer numbering
the slices.

Formally, the solution can be expressed as a function of a new set
of independent variables $(\sigma,\theta)$ according to
\begin{equation}
t=\sigma+\theta\tau,\label{STR}
\end{equation}
where $\sigma\in[0,\tau)$ is the \textit{pseudo-space} and $\theta=0,1.2,..$
the \textit{pseudo-time}. We remark that the values of the solution
are not affected by the use of the new variables, it is only a different
parametrization of the independent variable $t$ . The description
of the solution based on (\ref{STR}) has been named Spatio-Temporal
Representation (STR) \cite{Vogel1965,Arecchi1992,Willeboordse1992,Fischer1995}.
We remark that, since the STR is a re-organization of samples collected
sequentially in time, it can be applied as well to data from experiments
where it can provide very useful insights. 

As shown in Fig.\ref{fig:STR-example}, the propagation of structures
(cellular pattern) is evident in such representation, very similarly
to what observed e.g. in coupled map lattices and in 1D spatially
extended system such as the Complex-Ginzburg Landau (CGL) model.

The inspection of Fig.\ref{fig:STR-example}(b) also reveals the existence
of a \textit{drift}: the pattern in a slice (almost) reappears in
the successive slice shifted by some value $\delta$. This feature
is typical of DDS's and it is a consequence of the causality principle
(we will return to this point in Sec.~\ref{sec:phys-mean}). The
use of a slightly modified representation, with the interval size
$\tau'=\tau+\delta$ (Fig.~\ref{fig:STR-example}(c)) corresponds
to the correct choice of a comoving reference frame and allows to
obtain an almost vertical propagation. As seen in the following, the
value of the drift $\delta$ can be found by studying the autocorrelation
properties of the solution, namely, the highest peak of the autocorrelation
function is at $\tau+\delta$.

\subsection{Autocorrelation analysis\label{sub:ac}}

The STR described in Sec.~\ref{sub:str} can be justified by the
analysis of the normalized autocorrelation function

\begin{equation}
r(\xi)=\frac{\left\langle \bar{x}(t)\bar{x}(t-\xi)\right\rangle }{\sigma_{x}^{2}},
\end{equation}
where $\left\langle \cdot\right\rangle $ denotes time average, $\bar{x}=x-\left\langle x\right\rangle $,
and $\sigma_{x}^{2}=\left\langle \bar{x}^{2}\right\rangle $.

The function $r$ is a standard tool for extracting important features
of the dynamics from a time series. It is proven to be very useful
when the state of a system is known to a limited extent, e.g. in the
case of data obtained from numerical integration of complex models
or from experiments.

\begin{figure}
\centering{}\includegraphics[width=0.6\textwidth]{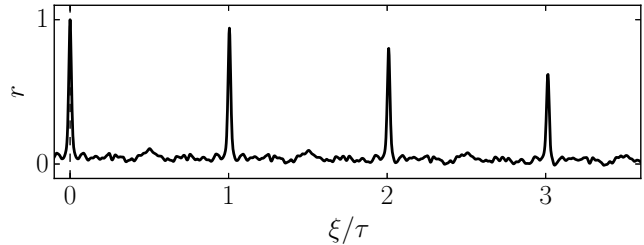}\caption{\label{fig:STR-ac}Autocorrelation function of a chaotic signal from
the integration of (\ref{eq:sl}) with $\tau=100$, $\mu=-1+i$, $\eta=1.9$,
and $\nu=-0.1-0.4i$.}
\end{figure}

The autocorrelation for a chaotic signal from Eq.~(\ref{eq:sl})
with large delay is plotted in Fig.~\ref{fig:STR-ac}, showing many
sharp peaks (revivals), nearly at the multiples of $\tau$. They indicate
the existence of temporal patterns almost repeating after a delay
time, resulting in the pseudo-time cellular structure in the STR.
As an effect of the chaotic dynamics, the height of the peaks is exponentially
decreasing, and their decay time (measured in units of the delay time)
gives an estimation of the average propagation length of the patterns
in the pseudo time. A more accurate inspection reveals that the peaks
are located at multiples of $\tau'=\tau+\delta$ and that the patterns
are therefore drifting in the STR as noted before. The decay time
$t_{c}$ of the first peak at $t=0$ (see Fig.~\ref{fig:STR-ac}),
is instead a measure of the average coherence length in the pseudo
space of such patterns. The above mentioned decays in the autocorrelation
can be interpreted as a signature of a (pseudo-) temporal and (pseudo-)
spatial disorder respectively. 

Even if other methods based e.g. on mutual information or fractal
dimension indicators are more powerful in the characterization of
the dynamics, the autocorrelation permits to identify the variables
to use for the STR. The multiple time scale analysis, as discussed
in the following sections, is a natural extension of this analysis
and allows to put on a rigorous ground the above considerations.

Since the system cannot respond (change significantly) on times shorter
than the coherence time, a rough estimation of the number of spatial
degrees of freedom can be heuristically evaluated by the number of
$t_{c}$ within the delay interval, i.e. $\rho=\tau/t_{c}$ \cite{LeBerre1987,Dorizzi1987,Lepri1993}.
Such a conjecture is supported by the observation that $t_{c}$ is
often independent of $\tau$ in the limit of large delays. This is
consistent with the reports that the attractor dimension scales linearly
with $\tau$ \cite{Farmer1982}. It should be noted that the existence
of a finite, non zero $t_{c}$ implies a limit in the maximal amount
of ``information processed'' by the delay system. In rigorous terms,
attractors in DDS's are shown to have finite dimension \cite{Mallet-Paret1976}. 

It is common to define ``long''-delay systems as those with $\rho\gg1$,
and it should be remarked that $t_{c}$ is not an ``instantaneous''
time-scale of the system (i.e., in the absence of feedback), but a
property of the delayed dynamics. On the other hand, $t_{c}$ is indeed
often close to the instantaneous time-scale in practice. 

\begin{figure}
\begin{centering}
\includegraphics[width=0.75\textwidth]{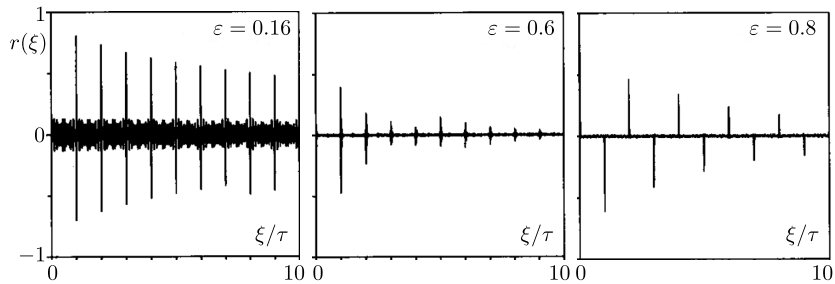}
\par\end{centering}

\caption{\label{fig:AC-log} Autocorrelation functions for the delayed logistic
map. (The figure is adapted with permission from Ref.~\cite{Lepri1993}.
Copyrighted by the American Physical Society.) }
\end{figure}

The sign of the peaks of the autocorrelation provide other important
information about the system, related to the phase of the feedback
signal. An example is shown in Fig.\ref{fig:AC-log}, in the case
of a delayed logistic map \cite{Lepri1993}. Negative revivals indicate
out-of-phase coherent structures contributing to the autocorrelation.
It should be noted that revivals are peaks in the \textit{envelope}
of the autocorrelation function: possible oscillations with negative
values at higher frequencies are specific of the signal itself but
do not carry information about the propagation. The difference is
very much the same as for the group and phase velocities of wave signals.
The case of negative revivals at the odd multiples of the delay time
\cite{Giacomelli1995,Javaloyes2015} can be represented with a suitable
STR, where the spatial cell is two- or more delays long. We will return
to this point in Sec.~\ref{sub:Stripes-and-square}.

The autocorrelation function can be shown to obey, in the case of
linear DDS's, the same equation as the original signal. Given the
system

\begin{equation}
\frac{dx}{dt}(t)=\mu x(t)+\eta x(t-\tau),\label{linear-model}
\end{equation}
the equation for the autocorrelation function \cite{Porte2014b} reads
as well

\begin{equation}
\frac{dr}{d\xi}(\xi)=\mu r(\xi)+\eta r(\xi-\tau).\label{linear-ac}
\end{equation}
 However, the initial condition for the evolution of the autocorrelation
$r(\xi),$ $-\tau<\xi\le0$ depends on the whole dynamics $x(t)$
and need not be the same as the initial condition $x(t),$ $-\tau<t\le0$
for the linear system (\ref{linear-model}). As a result, the exact
behavior of $r(\xi)$ and $x(t)$ can be different. Nevertheless,
Eq.~(\ref{linear-ac}) indicates that the values of the autocorrelation
in the following delay intervals, corresponding to separations larger
than $\tau$, are completely determined by the previous ones, and
this dependence has the same properties as the properties of the solution
itself, i.e. the STR holds for $r(\xi)$, the value of the drift is
the same, etc. 

The discussion above is rigorously valid for the linear model only.
However, one could expect that a similar behavior is observed in more
complicated situations. The autocorrelation has shown to be very useful
also in experiments \cite{Giacomelli2003,Porte2014b}; a STR description
for the autocorrelation can be derived from experimental data \cite{Giacomelli2003}.
In this case, the averaging process is very effective in improving
the signal-to-noise ratio and allows to obtain better results for,
e.g., an estimation of the drift. 

In setups with multiple feedback terms with different delays, the
structure of the peaks of the autocorrelation function can be more
complicated, reflecting the role of the multiple timescales involved.
A STR is possible in this case also, with a proper choice of the variables
and taking care of the drift over all components. We will discuss
this topic in Secs.~\ref{sub:normalform-2D} and \ref{sub:Spiral-defects-and}.

\subsection{The discrete case: maps with long delay\label{sub:maps}}

Although the main emphasis of this manuscript is on time-continuous
DDS, this section briefly reviews some results on the delayed maps.
Iterated maps is a large and important class of dynamical systems
with discrete time. The idea of a discrete dynamics with non-local
feedback emerged early in the discussions on numerical integration
of DDS's \cite{Farmer1982} and then they were studied by themselves
as simple but very general systems.

\begin{figure}
\begin{centering}
\includegraphics[width=0.6\columnwidth]{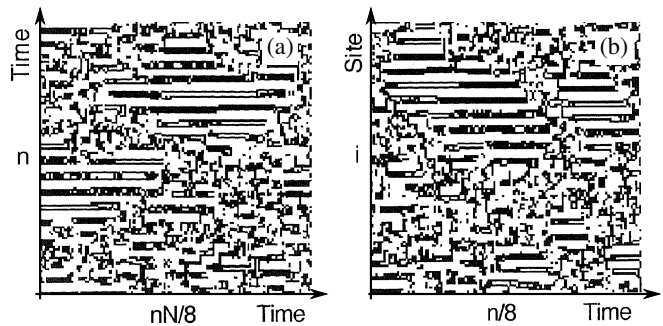}
\par\end{centering}

\caption{\label{fig:Time-time}Time-time diagram for the time delayed map (a)
and space-time diagram for the coupled-map lattice (b); every eighth
step in the $x$-direction was plotted. (From \cite{Willeboordse1992}:
Time-delayed map phenomenological equivalency with a coupled map lattice,
F.~H.~Willeboordse, Int. J. of Bifurcations \& Chaos 2, 721, Copyright
@ 1992 World Scientific Publishing Co., Inc.)}
\end{figure}

Willeboordse \cite{Willeboordse1992} studied a map with delay as
an alternative, algorithmic description of a Coupled Map Lattice (CML),
providing similar behaviors. Indeed, he proposed a method to derive
a delayed map from a CML, as a convenient method to obtain numerically
the same ``bulk'' dynamics. The mapping between the two systems
is a first example of pseudo spatio-temporal (in fact, called time-time)
representation (Fig.~\ref{fig:Time-time}). He noticed also that
the different boundary conditions may play an essential role in some
cases.

After the introduction of the STR, long delayed maps were first investigated
in \cite{Lepri1993,Lepri1994,Giacomelli1995}. The studies were motivated
both by the experiments in a $CO_{2}$ laser with delayed \cite{Arecchi1991}
and long delayed \cite{Arecchi1992} feedback on the losses, and from
the growing interest in CML's \cite{Kaneko1984-1,Kaneko1984-2,Kaneko1985}.

In \cite{Lepri1993} two classes of delayed maps are discussed. The
first (named class-I), is the discretization obtained e.g. by implementing
the forward Euler integration scheme of the model

\begin{equation}
\frac{dy}{dt}(t)=-y(t)+F(y(t-\tau))\label{eq:class-I}
\end{equation}
including the Mackey-Glass \cite{Mackey1977a} and Ikeda \cite{Ikeda1979}
delayed equations. Its peculiarity is that the local damping term
is strictly negative, with important effects on the type of the dynamics
in the large delay limit.

The second class has the form

\begin{equation}
y_{t+1}=(1-\varepsilon)f_{1}(y_{t})+\varepsilon f_{2}(y_{t-T}),\label{eq:class-II}
\end{equation}
which is not a discretized version of some differential equation,
but is more general, with some nonlinear functions $f_{12}$ and $\varepsilon$
measuring the relative weight of the delayed coupling. Here $T$ is
an integer delay time; in the case of class-I systems it can be obtained
from $T=\tau/t_{c}$ as discussed in Sec.~\ref{sub:ac}. Using both
the Bernoulli and logistic map nonlinear functions, the scaling of
Lyapunov spectrum, fractal dimension, and Kolmogorov-Sinai entropy
with the delay time is studied in \cite{Lepri1993} for different
values of the parameter $\varepsilon$.

For systems (\ref{eq:class-II}), the temporal evolution of small
perturbations $\delta y_{t}$ is described by the linearized map

\begin{equation}
\delta y_{t+1}=(1-\epsilon)f_{1}^{'}(y_{t})\delta y_{t}+\epsilon f_{2}^{'}(y_{t-T})\delta y_{t-T}.\label{eq:lin-map}
\end{equation}
Assuming a finite growth rate of $\delta y_{t}$, e.g. $\delta y_{t}\sim\mu^{t},$
$\left|\mu\right|>1$, in the limit $T\to\infty$, the retarded term
$\delta y_{t-T}\sim\mu^{-T}\delta y_{t}$ can be neglected. In this
case, the maximum Lyapunov exponent is given by

\begin{equation}
\lambda_{max}=\ln\left|\mu\right|=\left\langle \ln|f_{1}^{'}(y_{t})|\right\rangle +\ln(1-\varepsilon).
\end{equation}
This approach is self-consistent and holds only if $\lambda_{max}>0$.
The exponent was named \textit{anomalous}, to underline its different
behavior in the large delay limit with respect to the rest of the
spectrum. The anomalous exponent describes a fast decorrelation of
the dynamics on short time scales, implying that the STR cannot be
used: the information about the pattern in the previous delay time
is lost. It was also pointed out that class-I systems cannot have
an anomalous exponent, because of the negative instantaneous damping.
However, it could be found in more complex setups, such as models
with a larger number of coupled variables or experiments. The topics
has been increasingly studied in discrete as well in continuous systems
and in experiments \cite{Yanchuk2005a,Yanchuk2006,Yanchuk2009,Wolfrum2010,Yanchuk2010a,Flunkert2010,Heiligenthal2011,Lichtner2011,DHuys2013,Kinzel2013,Heiligenthal2013,Oliver2015,Juengling2015b,Juengling2015a},
and the terms \textit{strong chaos, }when the anomalous exponent exists,
and \textit{weak chaos} has been introduced, remarking the difference
between the two states. This point will be treated in details in Sec.~\ref{sec:LyapunovSpectrum}
(general properties of the spectrum) and Sec.~\ref{sub:Strong-and-weak}
(strong and weak chaos). 

Apart from the anomalous contribution, the Lyapunov spectrum has been
shown to scale with $1/T$ confirming the findings in differential-delay
equations \cite{Farmer1982,Ikeda1987}. From the knowledge of the
spectrum both the Kolmogorov-Sinai (or metric) entropy and the fractal
dimension can be calculated. This will be done in Sec.~\ref{sub:Entropy and KY}.
As a result, it is found that the dimension grows linearly with delay
$T$ (exactly as it was found in CML's) allowing to identify the delay
$T$ as an effective system size. The only remarkable exception is
the presence of an anomalous exponent, which upon rescaling is a diverging
quantity; the connection with CML's fails in this case.

The role of (pseudo) spatio-temporal chaos in determining nonequlibrium
statistical properties is reported in \cite{Lepri1994}, in the case
of a class-II type model based on logistic maps i.e. $f_{1}(x)=f_{2}(x)=\mu x(1-x)$
where $x\in[0,1]$. In this work, it is described and characterized
the appearance of a phase transition mediated by chaotic defects in
the STR, between a laminar and a turbulent regime with defects creation,
annihilation and diffusion. The measured scaling exponents at the
transition are in good agreement with those found in the CML's.

\begin{figure}
\begin{centering}
\includegraphics[scale=0.4]{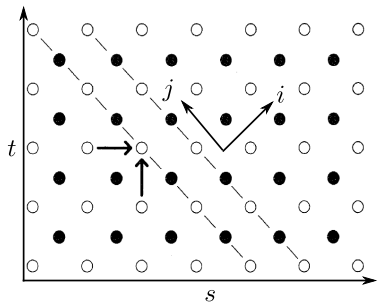}
\par\end{centering}

\caption{\label{fig:Giacomelli1995} Coupling scheme for the map (\ref{eq:dm}).
Coordinates $(s,t)$ correspond to the STR, while coordinates $(i,j)$
of the tilted reference frame allow to rewrite the map as an extended,
discrete system using two sub-lattices (empty and filled dots). (The
figure is adapted with permission from Ref. \cite{Giacomelli1995}.
Copyrighted by the American Physical Society.)}
\end{figure}

The statistical properties of the bidimensional patterns generated
by the dynamics might depend on the choice of the updating rule and/or
of the boundary conditions, even if the large aspect ratio (or delay)
limit is achieved. In the case of class-II models, an investigation
was reported in \cite{Giacomelli1995}. The evolution equation for
a generic delayed map $x_{n+1}=F(x_{n},x_{n+1-T})$ in the STR $n=s+tT$
can be written as 

\begin{equation}
y_{s}^{t+1}=F(y_{s-1}^{t+1},y_{s}^{t}),\label{eq:dm}
\end{equation}
with the boundary condition $y_{0}^{t+1}=y_{T}^{t}.$ The updating
rule is clearly sequential, since we must proceed from $s=1$ to $T$.
In the representation (\ref{eq:dm}) the coupling becomes local and
the system is in the class of 1D extended systems. Considering the
pattern generated by the rule (\ref{eq:dm}), a different coordinate
systems on it can be chosen by introducing a reference frame rotated
by $\pi/4$ (see Fig.~\ref{fig:Giacomelli1995}). Every point can
be parametrized with the integers $i=s+t$ and $j=s-t$. Denoting
the new variable as $z_{j}^{i}$, the map reads as 

\begin{equation}
z_{j}^{i+1}=F(z_{j-1}^{i},z_{j+1}^{i});
\end{equation}
the lack of the term $z_{j}^{i}$ in the coupling implies the existence
of two sub-lattices (depicted in Fig.~\ref{fig:Giacomelli1995} by
empty and, respectively, filled circles). This case appears analogous
to what found in CML's for $\varepsilon=1/2$; however, boundary conditions
and symmetries here are radically different. 

Considering three different rules to generate the same pattern, such
that the dynamical equations are the same but the boundary conditions
are different, the authors of \cite{Giacomelli1995} investigated
the behavior of the statistical indicators as a function of the coupling
parameter. They found that the Kolmogorv-Sinai entropy density (i.e.
the entropy divided by the delay) does not depend on the rule, and,
hence, it may be associated to the pattern. The Kaplan-Yorke dimension
density, as a consequence of the same property of the Lyapunov spectrum
is instead rule dependent, and, hence, the diverse generated patterns
show a different number of degrees of freedom.

The case of discrete, stochastic processes with two-delays is similarly
analyzed in \cite{Dahmen2008}. There, a time series evolving according
to a nonlocal update rule can be mapped onto a local process in two
dimensions with special time delayed boundary conditions. 

The transition to chaos synchronization in networks of iterated maps
in the limit of large delay times has been investigated in \cite{Flunkert2010,Zeeb2013}.
In particular, in \cite{Zeeb2013}, general arguments about the dynamics
of such a network predict a jump of the Kaplan-Yorke dimension of
the chaotic attractor when the network synchronizes. In addition,
the Kolmogorov entropy can show a discontinuous slope. This stays
in accordance with the fact, that the synchronization is related to
the bifurcation transition, at which the dynamics starts to be confined
to some low-dimensional synchronization manifold. As a result, the
Kaplan-Yorke and the correlation dimension jump to a lower value when
the network synchronizes.

\subsection{Long delay dynamics as composition of operators}

A formal decomposition of the class of models (\ref{eq:class-I})
was presented in \cite{Giacomelli1996}. There, it was pointed out
that the dynamics can be formally described as the action of two successive
operators on the solution in the segment $(0,\tau)$:

\begin{equation}
y_{d}\rightarrow z_{d}=F(y_{d}),\quad z_{d}\rightarrow y=\mathcal{L}z_{d}.
\end{equation}
The first operator is a nonlinear transformation at every point (i.e.,
it is a local operator in the pseudo space), playing the same role
of the nonlinear function in the CML's. The second is a linear operator,
combining the differential operator and the (pseudo) time evolution
from a delay to the successive one. The operator $\mathcal{L}$ can
be written in the Fourier space as

\begin{equation}
y(k)=\frac{z_{d}(k)}{1+ik},
\end{equation}
where $k$ is the spatial wavevector, and approximated up to the second
order by

\begin{equation}
y(k)\approx z_{d}(k)e^{-ik}e^{-k^{2}/2}.\label{eq:lin-op-approx-1-1}
\end{equation}

It is apparent from (\ref{eq:lin-op-approx-1-1}) that $e^{-ik}$is
responsible for the drift from a delay to the successive one, and
the Gaussian factor describes a diffusion process. Simulations showed
a good agreement of such an interpretation with a direct numerical
integration of the original model. The class (\ref{eq:class-I}) can
be thus described as an intermediate type between space-extended systems
(due to the action of drift and diffusion in the pseudo space) and
maps (iteration via a nonlinear local function).

\section{Lyapunov spectrum and other dynamic indicators \label{sec:LyapunovSpectrum}}

In this section, we discuss the Lyapunov spectrum (see e.g. \cite{Pikovsky2016})
and other important dynamical indicators used for the analysis of
delayed models. Particularly, we show how for long delays the spectrum
splits into a pseudo-continuous part and another, finite set of exponents.
This fact constitutes the theoretical foundation for the observation
that the long-delay systems incorporate the properties of both spatially
extended and finite-dimensional dynamical systems. The properties
of pseudo-continuous spectrum allow to provide a universal classification
of instabilities of long-delayed systems in analogy to the spatially-extended
systems, as discussed in Sec.~\ref{sub:Classification-of-instabilities}.
The case of several hierarchical time delays is studied in Sec.~\ref{sub:Spectrum-multiple-delays}.
In Sections~\ref{sub:Scaling-of-Floquet} and \ref{sub:ws}, the
Floquet and Lyapunov exponents (LE) of periodic and, respectively,
chaotic solutions are considered. Finally, we discuss in Sec.~\ref{sub:Entropy and KY}
the scaling of the entropy and Kaplan-Yorke dimensions of the attractors
in delayed systems.

\subsection{Spectrum of long-delay systems \label{spectrum}}

The dynamics in a close vicinity of any solution $s(t)$ of delay
system (\ref{general}) and its stability properties are determined
by the linearized system 
\begin{equation}
\frac{dx}{dt}(t)=A(t)x(t)+B(t)x(t-\tau),\label{eq:lin}
\end{equation}
where $A(t)=\partial_{1}F(s(t),s(t-\tau))$ and $B(t)=\partial_{2}F(s(t),s(t-\tau))$
are the derivatives (Jacobians) with respect to the first and the
second arguments, respectively. When the considered solution is stationary
$s(t)=s_{0}=\mbox{const}$, the matrices $A$ and $B$ are constant,
and the properties of (\ref{eq:lin}) are described by the solutions
of the characteristic equation 
\begin{equation}
\det\left[-\lambda I+A+Be^{-\lambda\tau}\right]=0.\label{eq:chareq}
\end{equation}
Here $I$ is the identity matrix. Equation (\ref{eq:chareq}) is a
quasi-polynomial, which, in general, possesses infinitely many complex
solutions. According to the general theory \cite{Hale1977}, they
all are bounded from the right, i.e. $\mbox{Re}\lambda<C$ and accumulate
solely at $-\infty$. Hence, only finitely many roots of (\ref{eq:chareq})
determine the stability. In particular, if there are roots with $\mbox{Re}\lambda>0$,
then the steady state $s_{0}$ is unstable, and if all roots have
$\mbox{Re}\lambda<0$, then it is asymptotically stable with the exponential
convergence rate $\exp((\mbox{Re}\lambda)_{\max}t)$. 

These basic properties of the spectrum already manifest some remarkable
features of the delay-systems. The infinite-dimensionality is expressed
by the infinite number of the characteristic roots $\lambda$. However,
only a finite number of them play a decisive role for the dynamics,
since the remaining part of the spectrum accumulates at $-\infty$
and corresponds just to an exponential (or sometimes even faster)
decay. This will also be reflected in the properties of periodic solutions,
as well as chaotic attractors. Roughly speaking, delay-equations is
an example of an infinite-dimensional system, that demonstrates finite-dimensional
phenomena in practice. The question is, how large the dimensionality
of the observed phenomena can be: as we will see in the following,
such a dimension can achieve any finite value and, in many cases it
grows linearly with the time delay $\tau$. 

\begin{figure}
\begin{centering}
\includegraphics[width=0.85\textwidth]{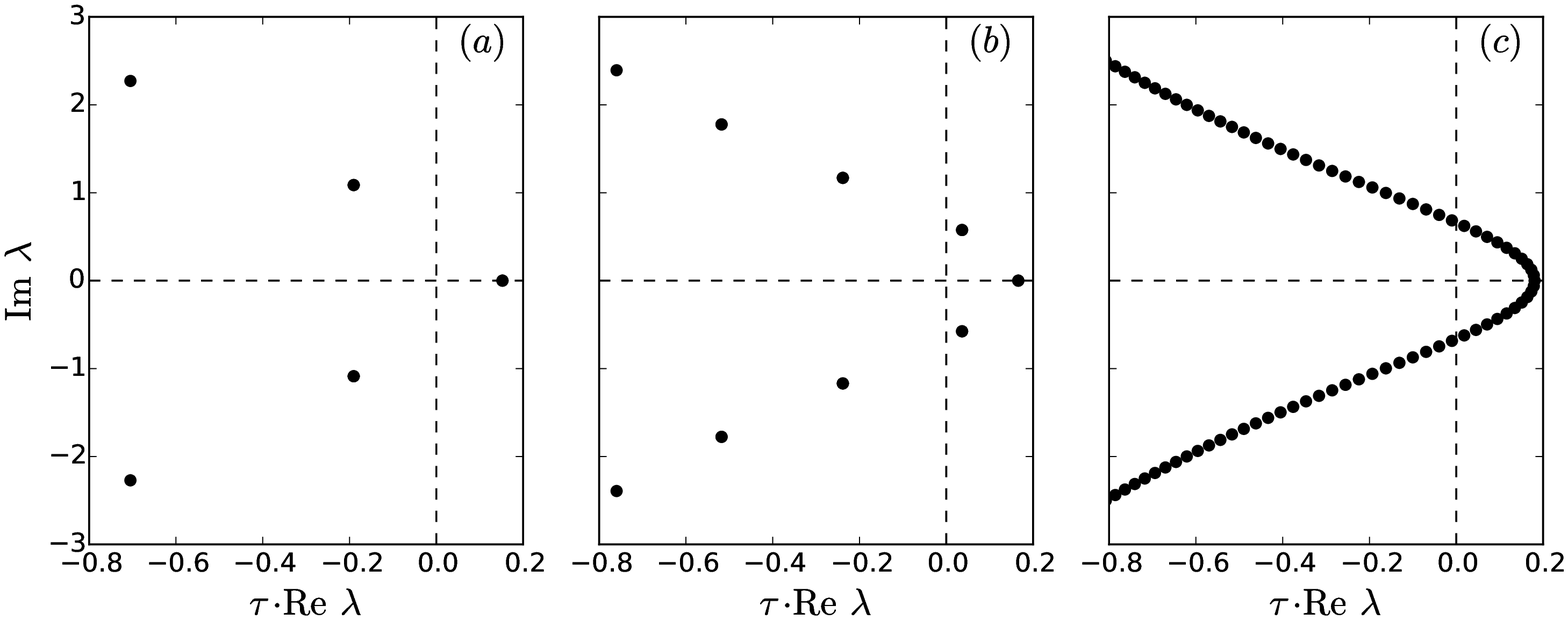}
\par\end{centering}

\caption{\label{fig:SimpleSpectrum}Spectrum of the scalar delay system $x'=ax(t)+bx(t-\tau)$
with $a=-1$ and $b=1.2$. As $\tau$ increases from (a) to (c), the
number of critical characteristic roots increases linearly and fill
the curve of the pseudo-continuous spectrum (\ref{eq:pcs}). Delay
values: $\tau=5$ (a), $\tau=10$ (b), and $\tau=100$ (c). Remark
that the real part is multiplied by $\tau$. }
\end{figure}

Coming back to the characteristic equation (\ref{eq:chareq}), an
important question is therefore how many multipliers are important
for the dynamics, or how many unstable multipliers may occur in the
system. May it happen that their number becomes extremely large so
that the system will start to exhibit essentially infinite-dimensional
properties? The answer to this question is affirmative, and such a
situation occurs when the time delay $\tau$ becomes large. Moreover,
the number of ''important'' characteristic roots grows linearly with
$\tau$. An illustration of this situation is shown in Fig.~\ref{fig:SimpleSpectrum},
where the roots of a scalar characteristic equation are shown for
three different values of time delay.

It is shown in \cite{Lichtner2011} that the major part of the roots
of (\ref{eq:chareq}) are converging to the following curves in the
complex plane 
\begin{equation}
\lambda=\frac{\gamma(\omega)}{\tau}+i\omega\label{eq:pcs}
\end{equation}
with the real parts scaled as $1/\tau$. By substituting (\ref{eq:pcs})
into Eq.~(\ref{eq:chareq}) and considering a leading order, it is
easy to find that 
\[
\gamma(\omega)=-\ln|Y(\omega)|,
\]
where $Y(\omega)$ is a root of the polynomial 
\[
\det\left[-i\omega+A+BY\right]=0.
\]
For instance, in the scalar case used in Fig.~\ref{fig:SimpleSpectrum}
\begin{equation}
\frac{dx}{dt}(t)=ax(t)+bx(t-\tau),\label{eq:scalar}
\end{equation}
this curve is given by
\begin{equation}
\gamma(\omega)=-\ln\left|(i\omega-a)/b\right|,\label{eq:pcs-scalar}
\end{equation}
and it has one maximum at $\omega=0$ independently of the choice
of the parameters $a$ and $b$. 

Although these curves are continuous, the spectrum, which converges
to them is discrete with the distances between the neighboring characteristic
roots scaling as $2\pi/\tau$. Moreover, the number of such curves
is proportional to the number of variables on which the delayed feedback
is acting. To be more specific, it is equal to the rank of the matrix
$B$. Further details on the spectrum for the steady states can be
found in \cite{Bellman1963,Hale1977}, and, for large delays in \cite{Giacomelli1998,Yanchuk2005,Wolfrum2010,DHuys2013,Yanchuk2015a}. 

Another part of the spectrum is unstable and it is close to the unstable
characteristic roots of the instantaneous system, i.e. the eigenvalues
$\lambda_{A}$ of the matrix $A$ with positive real parts. For long
delay, $\lambda_{A}$ approximates a characteristic root of the delayed
system and leads to the instability. Such spectrum is also called
strongly unstable \cite{Lichtner2011} or anomalous \cite{Giacomelli1998},
and it contains only a finite number of roots (or empty) that is equal
to the number of unstable eigenvalues of the matrix $A$. 

Summarizing, the spectrum of systems with long delays, with increasing
delay starts to resemble the spectrum of spatially-extended systems,
where the curves of the pseudo-continuous spectrum are analogous to
the so-called dispersion-relation curves \cite{Cross1993,Cross2009}.
On the other hand, the characteristic roots are discrete as in spatially-extended
systems on a bounded spatial domain.

\subsection{Classification of instabilities in systems with long delays \label{sub:Classification-of-instabilities}}

\begin{figure}
\begin{centering}
\includegraphics[width=0.85\textwidth]{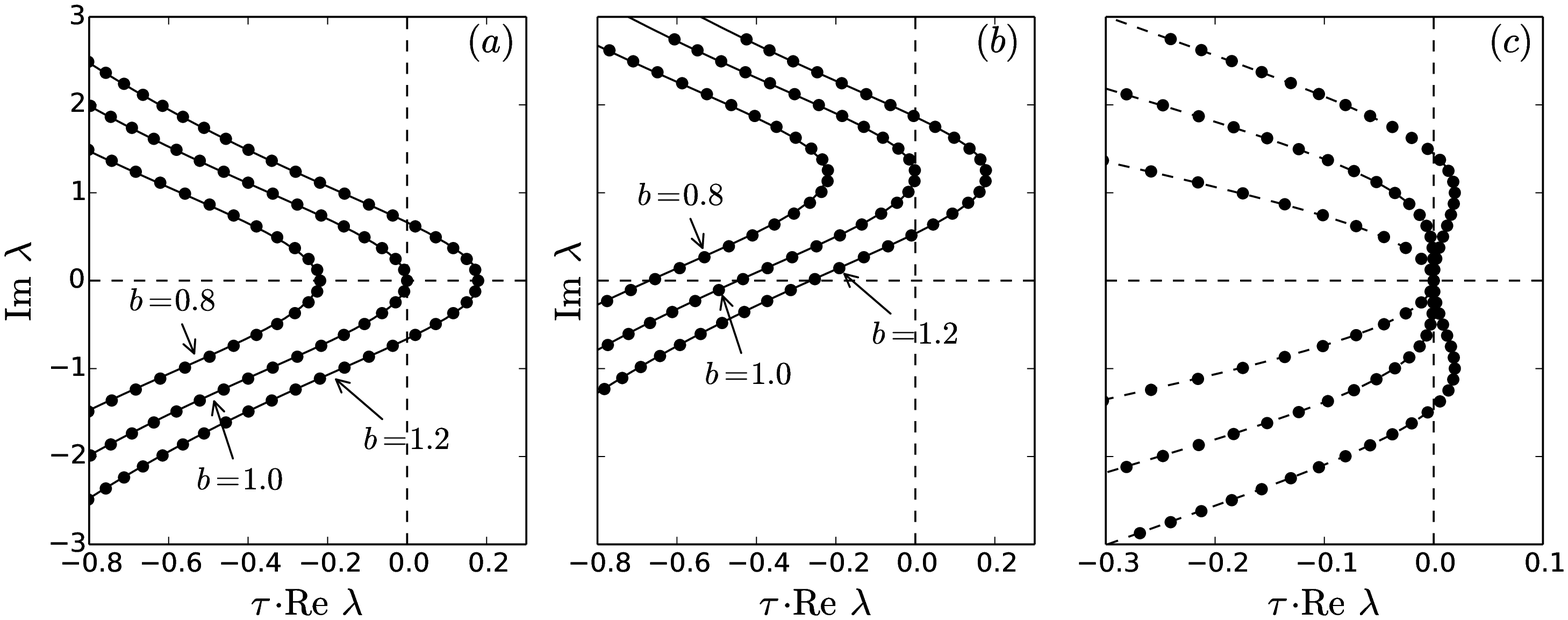}
\par\end{centering}

\caption{Classification of instabilities in large-delay systems. From left
to right: uniform, wave (Turing), and modulational. \label{fig:instabilities}}
\end{figure}

The destabilization in systems with long time delays can not be adequately
described by individual local bifurcations (e.g. Hopf) when a pair
of characteristic roots crosses the imaginary axis, but rather crossing
of the families of characteristic roots located along a branch of
the pseudo-continuous spectrum. The interplay of all Hopf bifurcations
produces essentially high-dimensional phenomena, which are typical
for spatially-extended systems. Figure~\ref{fig:instabilities} illustrates
the main (codimension-1) destabilizations that may occur \cite{Wolfrum2010}.
They can be classified into uniform, wave, and modulational, depending
on the way the pseudo-continuous spectrum crosses the imaginary axis.
This classification is inspired by the corresponding classification
existing in PDE systems \cite{Cross1993}. 

\emph{Uniform} instability occurs when the spectrum is tangent to
the imaginary axis at $\omega=0$, see Fig.~\ref{fig:instabilities}(a).
The generic crossing in this case is quadratic. The simplest model
where such an instability is present is the scalar equation (\ref{eq:scalar}).
In fact, this is the only instability that may occur for steady states
in the scalar case (\ref{eq:scalar}) with real coefficients, since
the pseudo-continuous spectrum has always the maximum at $\omega=0$
in this case, see \cite{Yanchuk2015a}. Fig.~\ref{fig:instabilities}(a)
was calculated for $a=-1$, $\tau=50$, and the $b$ parameters as
indicated in the figure. The dashed line shows the pseudo-continuous
spectrum line given by (\ref{eq:pcs}) and (\ref{eq:pcs-scalar}). 

\emph{Wave (or Turing}\footnote{It seems that there is still no common agreement about how to call
this instability. In some previous publications, also by the authors,
the name Turing instability was used. Here, we use the name ''wave
instability'', since it seems to be more appropriate for this case.}\emph{)} instability happens when the crossing takes place at $\omega\ne0$,
see Fig.~\ref{fig:instabilities}(b). For such an instability of
steady states to occur, at least two variables should be present in
the system. The simplest equation possessing the aforesaid instability
is the complex-valued, scalar equation (\ref{eq:scalar}) with complex
coefficients. Figure \ref{fig:instabilities}(b) was obtained for
$a=-1+i$ and $\tau=50$. 

\emph{Modulational} instability is generic for the destabilization
of systems with symmetry or periodic solutions, where a zero characteristic
root appears naturally and persists for all parameter values, see
Fig.~\ref{fig:instabilities}(c). Similarly to wave-instability,
several (i.e. more than one) variables are needed in order to observe
it. 

The above mentioned instabilities play often a fundamental role in
the appearance of various nonstationary patterns \cite{Wolfrum2006,Yanchuk2006,Yanchuk2010a,Wolfrum2010,Puzyrev2014,Yanchuk2015a,Puzyrev2016},
and, in the following sections we will mention them in the contexts
of emerging spatio-temporal phenomena.

\subsection{Spectrum in systems with multiple hierarchical time delays\uline{\label{sub:Spectrum-multiple-delays}}}

\begin{figure}
\begin{centering}
\includegraphics[width=0.8\textwidth]{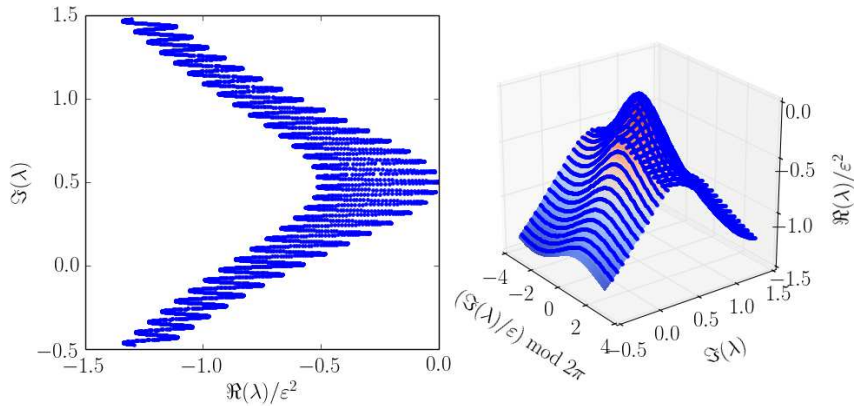}
\par\end{centering}

\caption{\label{fig:2dspectrum} Example of the spectrum for system $x'=ax(t)+bx(t-\tau)+cx(t-\tau_{2})$
with $a=-0.4+0.5i$, $b=0.1$, $c=0.3$, $\tau_{1}=100$, and $\tau_{2}=10^{4}$.
Points denote the characteristic roots computed numerically. The right
panel shows the surface of the 2D pseudo-continuous spectrum, which
is given analytically by the function (\ref{eq:gamma2D}), together
with the numerical points. }
\end{figure}

In the case of multiple hierarchical time delays $1\ll\tau_{1}\ll\tau_{2}\ll\cdots\ll\tau_{n}$,
the spectrum can also be divided into various components, which correspond
to each timescale $\tau_{i}$ \cite{Yanchuk,Yanchuk2015c,Yanchuk2014,DHuys2013}.
For instance, in the case of the system with two delays 
\[
x'=Ax(t)+B_{1}x(t-\tau_{1})+B_{2}x(t-\tau_{2}),
\]
the $\tau_{1}$-spectrum has still the form of a one-dimensional pseudo-continuous
spectrum. However, the destabilization in such systems is governed
generically by the $\tau_{2}$-spectrum, which scales as 
\[
\lambda=\frac{\gamma(\omega,\varphi)}{\tau_{2}}+i\omega,\quad\varphi=\left(\frac{2\pi\omega}{\tau_{1}}\right)\,\mbox{mod}\,2\pi,
\]
and it is asymptotically located on a two-dimensional surface $\gamma(\omega,\varphi)$,
which can be found explicitly as 
\begin{equation}
\gamma(\omega,\varphi)=-\ln|Y(\omega,\varphi)|,\,\,Y\,\mbox{satisfies}\,\det\left[-i\omega I+A+B_{1}e^{i\varphi}+B_{2}Y\right]=0.\label{eq:gamma2D}
\end{equation}
As a result, the destabilization is characterized by a family of characteristic
roots laying on a surface, just as in the case of a two-dimensional
dispersion relation in spatially-extended systems. Example of such
a destabilizing surface is shown in Fig.~\ref{fig:2dspectrum}. More
details can be found in \cite{Yanchuk,Yanchuk2015c,Yanchuk2014,DHuys2013},
where, in particular, the interplay between different types of spectra
is explained. 

When the number of delays is $k$, the destabilization is governed
by a $k$-dimensional surface \cite{Yanchuk}. 

As a result, in systems with hierarchical time delays, various spatiotemporal
phenomena can be observed, which are similar to those in spatially-extended
systems with higher number of spatial dimensions. Examples of such
phenomena, 2D spirals and defect turbulence, will be given in Sec.~\ref{sub:Spiral-defects-and}.

\subsection{Scaling of the Floquet exponents of periodic solutions \label{sub:Scaling-of-Floquet}}

The scaling properties of the steady states spectrum discussed in
the previous section, hold for periodic solutions as well \cite{Yanchuk2009,Sieber2013}.
Given a periodic solution $s(t)$ of system (\ref{general}), the
linearized equation around this periodic solution has the form (\ref{eq:lin})
with periodic matrices $A(t)$ and $B(t)$. The Floquet exponents
$\lambda$ measure the exponential growth of small perturbations from
this periodic solution, i.e. the growth rate of the solutions of the
linearization (\ref{eq:lin}). In particular, it is known \cite{Hale1977}
that for any Floquet exponent $\lambda$ there is a solution $x(t)$
of the linearized equation such that $x(t)=e^{\lambda t}p(t)$ with
periodic $p(t)$. The inverse is also true. 

As it is shown in \cite{Yanchuk2009,Sieber2013}, the spectrum of
Floquet exponents converges for large delays asymptotically to curves
with the same asymptotic properties (\ref{eq:pcs}) found for steady
states, with real parts scaling as $1/\tau$. For periodic solutions,
an additional consideration is needed to explain the meaning of the
large delay approximation. In contrast to steady states, which are
independent on time delay and solve the equation $F(s,s)=0$, the
periodic solutions $s(t)=s(t+T)$ are changing their shape and period
$T$ when the delay $\tau$ is changed. However, due to the periodicity
of $s(t)$, it holds $s'(t)=F(s(t),s(t-\tau_{0}))=F(s(t),s(t-kT-\tau_{0}))$,
and, hence the same periodic solution $s(t)$ exists for all delay
values $\tau=\tau_{0}+kT$, $k\in\mathbb{Z}$. As a consequence, one
can consider the stability of the solution $s(t)$ as the delay increases
by steps of $T$. 

From a practical point of view, the pseudo-continuous spectrum does
provide a good approximation for the spectrum of Floquet exponents
when the period of the periodic solution is much smaller than the
time delay.

\subsection{Scaling of Lyapunov exponents \label{sub:ws} }

As discussed above, the basic property of the spectrum of systems
with long delays is the existence of families of LEs (resp. eigenvalues
or Floquet exponents) that are asymptotically continuous. In the case
of chaotic systems, this implies the appearance of so called weak
chaos characterized by a set of unstable LEs, which scale as $1/\tau$,
and their number is proportional to $\tau$. As a result, extensive
chaos, or, hyperchaos with large number of unstable dimensions arises
\cite{Lepri1993,Heiligenthal2011,DHuys2013,Kinzel2013}. When instead
the largest LE does not scale with $\tau$, i.e. it is of order 1,
strong chaos regime sets in. 

The phenomena specifically associated with weak and strong chaos will
be discussed later in Sec.~\ref{sub:Strong-and-weak}. Here, we focus
on the reasons for the appearance of the corresponding scaling of
LEs. Formally, the LEs are defined as the growth rate of the perturbations
for the linearized equation (\ref{eq:lin}), where $A(t)$ and $B(t)$
are possibly irregular functions defined as Jacobians evaluated at
a given orbit $s(t)$ within a chaotic attractor. The LEs computed
from the truncated ODE system 
\begin{equation}
\frac{dx}{dt}(t)=A(t)x(t)\label{eq:trunc}
\end{equation}
 play a special role. Note here that $A(t)=\partial_{1}F(s(t),s(t-\tau))$
is still evaluated at the solution $s(t)$ of the full nonlinear delayed
system (\ref{general}). In \cite{Heiligenthal2011,Heiligenthal2013,DHuys2013,Kinzel2013},
they are called instantaneous or sub-Lyapunov exponents, or in \cite{Lepri1993},
anomalous LE's. 

In particular, when there is a positive sub-Lyapunov exponent $\lambda_{s}>0$,
it can be shown that there is a LE $\lambda$ of the original delay
system that is close to $\lambda_{s}$ for long time delays. In this
case, the maximal LE does not scale with time delay, and, hence, the
perturbations $x(t)$ grow on a timescale $t\sim1/\lambda_{s}$, which
is smaller than the delay interval. This situation is called strong
chaos, see Sec.~\ref{sub:Strong-and-weak}, and the corresponding
LE is called strongly unstable. It is straightforward to see the analogy
between such a LE and the strongly unstable eigenvalues or Floquet
multipliers for the case of steady states and periodic solutions mentioned
in Secs.~\ref{spectrum} and \ref{sub:Scaling-of-Floquet}.

In the case when all sub-Lyapunov exponents are negative, the whole
spectrum of LE's scales as $1/\tau$. This can be supported by the
following reasoning \cite{Heiligenthal2011} applied to the linearized
equation (\ref{eq:lin}). Let $t=\sigma+\tau j,$ where $0\le\sigma\le\tau$
and $j=\left\lfloor t/\tau\right\rfloor $ is the number of full delays
$\tau$ within the interval of the length $t$. Denote $\xi_{j}(\sigma):=x(\sigma+\tau j).$
The linear equation (\ref{eq:lin}) can be written with respect to
$\xi_{j}$ as follows 
\begin{equation}
\frac{d\xi_{j}}{d\sigma}(\sigma)=A_{j}(\sigma)\xi_{j}(\sigma)+B_{j}(\sigma)\xi_{j-1}(\sigma),\label{eq:linresc}
\end{equation}
where $A_{j}(\sigma)=A(\sigma+\tau j)$ and $B_{j}(\sigma)=B(\sigma+\tau j)$.
Let $X(t,t')$ be the fundamental solution of the truncated system
(\ref{eq:trunc}). Since we assume that all LE's of the truncated
system are negative, we have that this fundamental solution is exponentially
decreasing as $\|X(t,t')\|\le Me^{-\lambda_{0}(t-t')}$ with some
positive constants $M$ and $\lambda_{0}$. Using the fundamental
solution $X$, the Eq.~(\ref{eq:linresc}) can be solved with respect
to $\xi_{j}(t)$ using the variation of constants formula 
\begin{equation}
\xi_{j}(\sigma)=X_{j}(\sigma,0)\xi_{j-1}(\tau)+\int_{0}^{\sigma}X_{j}(\sigma,\sigma')B_{j}(\sigma')\xi_{j-1}(\sigma')d\sigma',\label{eq:var}
\end{equation}
where $X_{j}(\sigma,\sigma'):=X(\sigma+\tau j,\sigma'+\tau j)$. From
Eq.~(\ref{eq:var}) and using the exponential decrease of $X$, one
can obtain the following estimate 
\[
\left|\xi_{j}\right|_{C}\le L_{j}\left|\xi_{j-1}\right|_{C}\le(L_{j}\cdot L_{j-1}\cdot\cdots L_{1})\left|\xi_{0}\right|_{C},
\]
where $\left|\xi_{j}\right|_{C}:=\sup_{\sigma}\left|\xi_{j}(\sigma)\right|$
is the maximum of absolute value of the function $\xi_{j}(\sigma)$
on a $j$-th delay interval. In fact, the constants $L_{j}$ are estimations
of the upper bound of the absolute value of the fundamental matrix
$\left|X_{j}(\sigma,\sigma')\right|$ of the truncated system on $j$-th
delay interval (all possible $\sigma$ and $\sigma'$). Assuming also
that the properties of the chaotic attractor are in some sense ''homogeneous''
such that there is a bound $L$ for all constants $L_{j}$, and the
estimate 
\[
\left|\xi_{j}\right|_{C}\le L^{j}\left|\xi_{0}\right|_{C}
\]
holds, we obtain 
\[
\left|x(t)\right|=\left|x(\sigma+\tau j)\right|\le\left|\xi_{j}\right|_{C}\le L^{j}\left|\xi_{0}\right|_{C}\le L^{t/\tau}\left|\xi_{0}\right|_{C}=e^{t\ln L/\tau}\cdot\mbox{const}.
\]
The obtained estimate shows that the exponential grows rate of $x(t)$
and, hence, the LE, scales at most as $1/\tau$ for the case, when
no strongly unstable LEs are present. We remark that the ''homogeneity''
assumption and the existence of a uniform bound for $L_{j}$ can be
made exact for the case of scalar DDS (\ref{general}), see \cite{Yanchuk2015a}.
For the vector case, the above reasoning seems to be plausible; however
a rigorous proof is still missing. 

We should also mention earlier works on LEs for DDEs \cite{Farmer1982,LeBerre1986,LeBerre1987},
where, in particular, the above scaling was observed.

\subsection{Entropy and Kaplan-Yorke dimension \label{sub:Entropy and KY}}

In this section, we shortly mention the consequences which follow
directly from the scaling properties of the Lyapunov spectrum. We
assume that the Pesin entropy formula for the metric entropy holds
\cite{Pesin1997} 

\begin{equation}
h_{\tau}=\sum_{\lambda_{i}>0}\lambda_{i},\label{eq:ks}
\end{equation}
where the sum is taken over all positive LEs. The entropy is the result
of the total asymptotic exponential rate of the expansion. In the
case of DDS with long delays, the sum can be splitted into two terms.
The first is for the strongly unstable spectrum (if it exists), and
the second for the positive part of the pseudo-continuous spectrum
\[
h_{\tau}=h_{su}+h_{pc}=\sum_{\mbox{str. unst.}}\lambda_{s,j}+\sum_{\begin{array}{c}
\lambda_{i}>0\\
\mbox{pseudo-cont.}
\end{array}}\lambda_{i}.
\]
Here we assumed a generic situation when no critical spectrum occurs,
i.e. there is no zero sub-Lyapunov exponent. Taking into account the
scaling properties of each part of the spectrum, we see that for a
long delay the first part $h_{su}\to h_{su}^{*}=\mbox{const}$ for
$t\to\infty$, and the second part 
\[
h_{pc}\to\sum\frac{\gamma_{j}}{\tau}\approx\int_{[0,\Omega_{+}]}\gamma(\omega)d\omega=h_{pc}^{*},
\]
converges also to a constant, that equals to the integral of the (rescaled)
asymptotically continuous spectrum on the range $[0,\Omega_{+}]$
where this spectrum is positive. The rescaled spectrum $\gamma_{j}$
should be considered as a function of the variable $\omega=j/\tau$.
Despite of the fact that the number of unstable LEs grows linearly
with $\tau$, their magnitude scales as $1/\tau$. As a result, this
integral converges to a constant when $t\to\infty$.

As a result, with $\tau\to\infty$ the entropy converges to a constant
value 
\[
h_{\tau}\to h_{\infty}=h_{su}^{*}+h_{pc}^{*},
\]
which is composed of two contributions, one from the strongly unstable
spectrum $h_{su}^{*}$, in the case of strong chaos (see Sec.~\ref{sub:Strong-and-weak}),
and another one ($h_{pc}^{*}$) for the contribution of the pseudo-continuous
spectrum. Here the duality of the DDS as a bridge between spatially-extended
finite-dimensional systems reveals again. 

The Lyapunov dimension of an attractor can be evaluated using the
well-known Kaplan-Yorke formula \cite{Kaplan1979}

\begin{equation}
d_{KY}=k+\frac{1}{|\lambda_{k+1}|}\sum_{i=1}^{k}\lambda_{i}.\label{eq:dky}
\end{equation}
The integer $k$ is the maximum value of $l$ such that the sum $\lambda_{1}+\cdots+\lambda_{l}$
is positive. The ordering $\lambda_{1}\ge\lambda_{2}\ge\cdots$ is
assumed. Here again, the sum $\sum_{i=1}^{k}\lambda_{i}$ can be splitted
into two parts: $h_{su}$ (it is zero when no strong spectrum is present)
and the sum over the interval $[0,\Omega_{0}]$ such that 
\begin{equation}
h_{su}+\int_{[0,\Omega_{0}]}\gamma(\omega)d\omega=0.\label{eq:ky}
\end{equation}
The value of $k$ is then given by $\left\lfloor \Omega_{0}\tau\right\rfloor $,
and linearly grows with $\tau$. Taking into account that the LEs
are at the points $i/\tau$, $i=1,2,\dots,$ we see that 
\[
\sum_{i=1}^{k}\lambda_{i}\approx h_{su}+\int_{[0,\Omega_{0}]}\gamma(\omega)d\omega+\left(\frac{\Omega_{0}}{\tau}-\left\lfloor \frac{\Omega_{0}}{\tau}\right\rfloor \right)\gamma(\Omega_{0})
\]
Since $\lambda_{k+1}\approx\frac{\gamma(\Omega_{0})}{\tau}$, we finally
obtain for the Kaplan-Yorke dimension the scaling 
\[
d_{KY}\approx\Omega_{0}\tau
\]
i.e. it grows linearly with the time delay, and $d_{KY}/\tau\to\Omega_{0}=\mbox{const}$
for $\tau\to\infty$.

\subsection{Growth of the number of coexisting periodic solutions with delay}

\begin{figure}
\centering{}\includegraphics[width=1\textwidth]{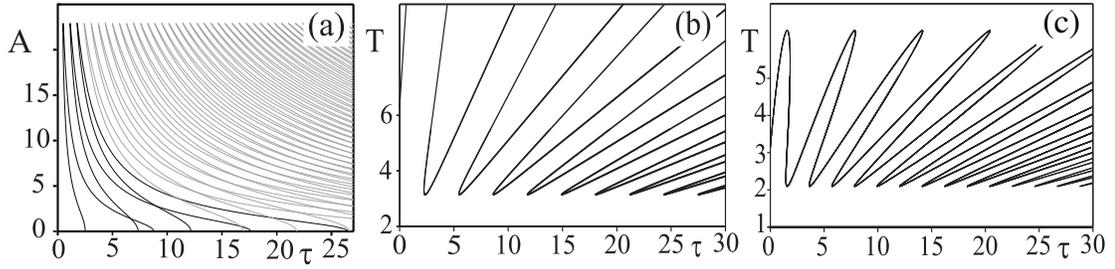}\caption{\label{fig:reappearance}Typical ``reappearance'' of the branches
of periodic solutions versus time delay $\tau$. (a) Duffing oscillator
(\ref{eq:duffing}) with delayed feedback; the amplitude $A$ of the
periodic solutions versus time delay $\tau$ is shown; parameters
are $a=0.5$, \textbf{$b=0.6$, }and\textbf{ $d=0.06$}. (b)-(c) Oscillator
(\ref{eq:sl}) with parameters $\nu=-1$, $\eta=1$, and $\mu=2+i$
for (b), $\mu=2+2i$ for (c), respectively.}
\end{figure}

An important complexity measure of a dynamical system is provided
by the number of coexisting periodic solutions. It is shown in \cite{Yanchuk2009}
that DDS (\ref{general}) generically possess coexisting periodic
solutions, the number of which grows linearly with $\tau$. Typical
examples are shown in Fig.~\ref{fig:reappearance}, where the branches
of periodic solutions are shown versus time delay. In particular,
Fig.~\ref{fig:reappearance}(a) corresponds to the Duffing oscillator
with delayed feedback 
\begin{equation}
x''(t)+dx'(t)+ax(t)+x^{3}(t)+b\left(x(t)-x(t-\tau)\right)=0,\label{eq:duffing}
\end{equation}
see the figure caption for the parameter values. One can observe how
the branch is created in Hopf bifurcations, and how it ``reappears''
with a modified shape for larger delays. Similarly, Figs.~\ref{fig:reappearance}(b)
and (c) show branches of periodic solutions for the Stuart-landau
oscillator (\ref{eq:sl}). 

In fact, such a reappearance is shown to be generic for systems with
time delay \cite{Yanchuk2009}. It is a consequence of an elementary
property of DDS such that any periodic solution $x(t)$ with a period
$T(\tau)$, which exists for a time delay $\tau$ also exists (reappears)
for the time delay $\tau+kT(\tau)$ with $k$ being an arbitrary integer
number. Moreover, the reappearing branches change their shape and
stretch with the increasing of the delay, thus overlapping with each
other, leading to the multistability. A proper counting of the number
of such overlapping branches gives the following estimate for the
number of coexisting periodic solutions 
\begin{equation}
N\approx\kappa\tau=\left(\frac{1}{T_{\min}}-\frac{1}{T_{\max}}\right)\tau,\label{eq:per}
\end{equation}
where $T_{\min}$ and $T_{\max}$ are the maximal and minimal period
along each ``elementary'' branch. In the case when the period is
unbounded, as in Fig.~\ref{fig:reappearance}(b), this formula reduces
to 
\begin{equation}
N\approx\kappa_{2}\tau=\frac{1}{T_{\min}}\tau.\label{eq:growthcase2}
\end{equation}
More details on the reappearance, the stability properties of the
branches, as well as on some associated bifurcation properties (note
a sequence of fold or Hopf bifurcations) can be found in \cite{Yanchuk2009}.
Notice also that there may exist few primary branches, which cannot
be mapped one onto another by the transformation $\tau\to\tau+kT(\tau)$.
In this case, each branch will reappear increasing the delay time.
The growth rate $\kappa$ is here a superposition of the corresponding
rates. Generally speaking, Eq.~(\ref{eq:per}) gives a lower bound
for the number of periodic orbits. 

For DDS with an additional $S^{1}$ symmetry, e.g. the Stuart-Landau
system, one can even show that the number of relative periodic solutions
(called sometimes modulated waves, i.e. the solutions of the form
$z(t)=e^{i\omega t}A(\beta t)$ with periodic $A(t)$) grows even
faster, as $\tau^{2}$ \cite{YanchukSieber2013}.

\section{Multiple time-scale analysis and normal forms \label{sec:math}}

In the previous sections we have shown that long-delayed systems are
characterized by a natural separation between different timescales.
This can be, in particular visualized by the space-time representation.
A more quantitative analysis, employing the autocorrelation function
as discussed in Sec. \ref{sub:ac} can furthermore estimate the degree
of coherence along such scales. In this section, we present and discuss
the applications of multiple timescale analysis methods to different
classes of long-delayed systems.

\subsection{\label{sub:Reduction-to-parabolic}Reduction to normal form close
to destabilization }

We start by briefly discussing how the Ginzburg-Landau equation can
be derived as a normal form close to a destabilization of a system
with long time delays. 

The reason why a parabolic partial differential equation (PDE) appears,
instead of a standard, low-dimensional normal form as in the case
of bifurcations in ODEs, becomes apparent when taking into account
the shape of the spectrum of long-delay systems. As described in Sec.~\ref{spectrum},
the spectrum approaches asymptotically a set of continuous curves
in the appropriately rescaled complex plane $(\tau\mbox{Re}\lambda,\mbox{Im}\lambda)$,
see e.g. Fig.~\ref{fig:instabilities}. As a consequence, the destabilization
is governed by the crossing of the curves of the imaginary axis, similarly
to the case of parabolic PDEs, where such curves are called dispersion
relations. In the case of PDEs, the normal forms at some destabilizations
are Ginzburg-Landau equations \cite{Cross1993,Cross2009}, and the
normal form equations depend only on the type of the destabilization
and not on the details of the original equations. 

It turns out that such normal forms can be derived for delay systems
as well. However, while the spatial coordinates in the PDE case are
naturally defined, in the delay case the role of the spatial coordinates
in the normal form is played by different scales of the time $\varepsilon t$,
$\varepsilon^{2}t$, etc.

\subsection{Uniform instability\label{sub:Normal-uniform}}

In this section, we give the main ideas for the derivation of the
normal form using the simple example of system (\ref{general}) with
a scalar variable $x(t)$. A more detailed derivation with a rigorous
proof can be found in \cite{Yanchuk2015a}. 

The main assumptions at the basis of the derivation are the following:
\\
(i) There is a steady state $x^{*}$, which can be set to zero by
a coordinate transformation.\\
(ii) The time delay is large $\tau\gg1$. We exclude also the existence
of other large parameters, or, equivalently, small parameters that
are on the order of $\varepsilon=1/\tau$ or smaller. \\
(iii) The steady state is close to a destabilization, i.e. the parameter
values are such that the spectral curve is crossing or about to cross
the imaginary axis, as depicted in Fig.~\ref{fig:instabilities}(a).
Since the pseudo-continuous spectrum (\ref{eq:pcs}) is given by the
function $\gamma(\omega)=-\ln\left|(i\omega-a)/b\right|$ (shown in
Sec.~\ref{spectrum}), the destabilization takes place for $a<0$
and $|a|=|b|$. Here $a$ and $b$ are the derivatives of the right
hand side with respect to the non-delayed, and delayed argument at
zero, respectively. Hence, the parameters of the system should be
close to this condition for the normal form to approximate the dynamics.
\\
(iv) Finally, we assume that the nonlinearity is of the third order
to exclude the case of transcritical bifurcation.

A general scalar delayed system, which satisfies the above conditions
(i)-(iv) can be formally written as 
\begin{equation}
x'(t)=\left(a+a_{1}\varepsilon^{2}\right)x(t)+(\pm a+b_{1}\varepsilon^{2})x(t-\tau)+f(x(t),x(t-\tau),\varepsilon^{2}),\label{eq:ddescalarclose}
\end{equation}
where the nonlinear function $f$ is of the third order in $x(t)$
and $x(t-\tau)$. We also point out that the perturbations $a_{1}\varepsilon^{2}$
and $b_{1}\varepsilon^{2}$ are of order $\varepsilon^{2}$, since
the amplitude of the solution emerging in the bifurcation scales as
$\varepsilon$ and, thus it will have a square root dependence on
the perturbation, similarly to the case of the Hopf or pitchfork bifurcation
\cite{Kuznetsov1995}. 

Substituting now the general multiscale Ansatz 
\[
x(t)=\varepsilon u_{1}(\varepsilon t,\varepsilon^{2}t,\varepsilon^{3}t)+\varepsilon^{2}u_{2}(\varepsilon t,\varepsilon^{2}t,\varepsilon^{3}t)+\cdots
\]
into Eq.~(\ref{eq:ddescalarclose}) and considering separately equations
obtained for the terms with different orders of $\varepsilon$ up
to the order $\varepsilon^{3}$ (solvability conditions, excluding
secular terms), one obtains the following result: 
\[
x(t)\approx\varepsilon A\left(\varepsilon t+\frac{1}{a}\varepsilon^{2}t,\varepsilon^{3}t\right),
\]
where the function $A(x,\theta)$ of two variables satisfies the PDE
\begin{equation}
\partial_{\theta}A\left(x,\theta\right)=\frac{1}{2a^{2}}\partial_{x}^{2}A\left(x,\theta\right)+\frac{1}{a^{2}}\partial_{x}A\left(x,\theta\right)-\frac{\left(a_{1}\mp b_{1}\right)}{a}A\left(x,\theta\right)-\frac{\varsigma}{a}A\left(x,\theta\right)^{3}\label{eq:NF}
\end{equation}
with the boundary condition 
\begin{equation}
A\left(x,\theta\right)=\mp A\left(x-1,\theta\right).\label{eq:NFBC}
\end{equation}
Here $\varsigma=\frac{1}{6}\left(f_{111}\mp3f_{112}+3f_{122}\mp f_{222}\right)$,
and the sub-indexing of $f$ denotes the partial derivatives with
respect to the corresponding variables. 

The variable $x=\varepsilon t+\frac{1}{a}\varepsilon^{2}t$ is called
pseudo-space and $\theta=\varepsilon^{3}t$ pseudo-time. Notice that
in such a way the introduced pseudo-time and pseudo-space variables
are in agreement with the geometric spatio-temporal description presented
in Sec.~\ref{sub:str}. Indeed, the spatial domain $0\le x\le1$
corresponds to an interval of $0\le t\le\frac{\tau}{(1+\varepsilon/a)}$,
which is equal to $\tau$ apart for the small correction due to the
drift. 

\begin{figure}
\centering{}\includegraphics[width=0.5\textwidth]{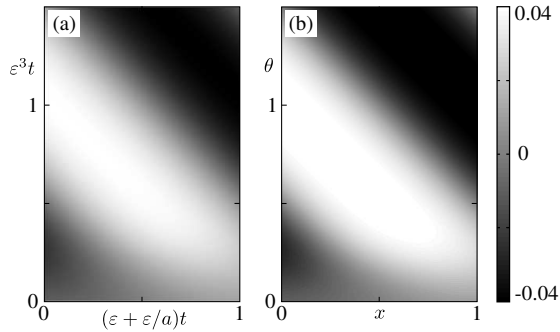}\caption{\label{fig:uniform-str}(a) Evolution for DDS (\ref{eq:ex}) in the
rescaled coordinates $x=\left(\varepsilon+\varepsilon^{2}/a\right)t$,
$\theta=\varepsilon^{3}t$. The magnitude of the solution $y(t)$
is shown by different scale of gray. (b) Evolution of the corresponding
solution of the related amplitude equation (\ref{eq:ex-ae}). }
\end{figure}

Figure~\ref{fig:uniform-str} shows an example for the scalar delay
equation 
\begin{equation}
y'(t)=-0.999y(t)-y(t-\tau)-0.5y^{3}(t)\label{eq:ex}
\end{equation}
with $\tau=100$. The solution is plotted in Fig.~\ref{fig:uniform-str}(a)
using the spatiotemporal coordinates $x=\left(\varepsilon+\varepsilon^{2}/a\right)t$
and $\theta=\varepsilon^{3}t$. Such a plot corresponds to the spatio-temporal
representation discussed in Sec.~\ref{sub:str}. Figure \ref{fig:uniform-str}(b)
shows the corresponding solution of the related to (\ref{eq:ex})
normal form 
\begin{equation}
\partial_{\theta}A=\frac{1}{2}\partial_{xx}A+\partial_{x}A+10A-\frac{1}{2}A^{3},\quad A\left(x,\theta\right)=-A\left(x+1,\theta\right).\label{eq:ex-ae}
\end{equation}
 As seen in the figure, the similarity between the two solutions is
a very high. 

Further details on the derivation of the normal form (\ref{eq:NF})\textendash (\ref{eq:NFBC})
as well as a rigorous error estimate can be found in \cite{Yanchuk2015a}.

\subsection{Wave instability in 1D \uline{\label{sub:Turing-instability-1D}}}

The wave destabilization takes place when the pseudo-continuous spectrum
is crossing the imaginary axis at a non-zero frequency (see Sec.~\ref{sub:Classification-of-instabilities}
and Fig.~\ref{fig:instabilities}(b)). Such a situation cannot occur
in the scalar case, since the pseudo-continuous spectrum there has
a unique maximum at $\omega=0$. In fact, at least two variables are
necessary to observe the wave instability. 

The simplest example of such a system is the scalar equation, with
a complex variable $z$ \cite{Wolfrum2006} 
\begin{equation}
\frac{dz}{dt}(t)=(\alpha+i\beta)z(t)-z(t)|z(t)|^{2}+z(t-\tau),\label{eq:PC-eckhaus}
\end{equation}
where $\alpha$ and $\beta$ are real parameters. The pseudo-continuous
spectrum for this system is given by the function 
\[
\gamma(\omega)=-\frac{1}{2}\ln\left[\alpha^{2}+(\omega-\beta)^{2}\right],
\]
which has a global maximum at $\omega=\beta$; hence, the wave destabilization
occurs when $\alpha$ becomes positive and $\beta\ne0$, see Fig.~\ref{fig:instabilities}(b).
Using a multiscale approach similar to that described in Sec.~\ref{sub:Normal-uniform},
the following normal form in the vicinity of this bifurcation can
be derived \cite{Wolfrum2006}: 
\begin{equation}
\partial_{\theta}A=\frac{1}{2}\partial_{x}^{2}A+\partial_{x}A+\mu A-A\left|A\right|^{2},\label{eq:cgl}
\end{equation}
equipped with the boundary condition $A(x,\theta)=e^{i\varphi}A(x-1,\theta)$
with some phase $\varphi$. The relation between the solution $z(t)$
of the delay equation and $A(x,\theta)$ from the normal form is 
\[
z(t)\approx\varepsilon e^{i\beta t}A(\varepsilon t-\varepsilon^{2}t,\varepsilon^{3}t).
\]
In contrast to the case of the uniform instability, the amplitude
$A(x,\theta)$ is complex-valued and it is determined by the complex
Ginzburg-Landau equation (\ref{eq:cgl}); it can therefore exhibit
more complicated dynamical phenomena than the real normal form (\ref{eq:NF}).
Moreover, the amplitude $A$ is now modulating the periodic ''signal''
$e^{i\beta t}$. One of the effects observed in such systems is the
Eckhaus phenomenon, which will be discussed in Sec.~\ref{sub:Eckhaus-instability}.

\subsection{Wave instability in 2D \uline{\label{sub:normalform-2D}}}

A paradigmatic model describing the wave instability in two dimensions
is provided by the equation similar to (\ref{eq:PC-eckhaus}) but
with two time delays, which are acting on different time scales \cite{Yanchuk2014,Yanchuk2015c}:
\begin{equation}
\frac{dz}{dt}(t)=az(t)+bz(t-\tau_{1})+cz(t-\tau_{2})+dz(t)|z(t)|^{2},\label{eq:2D}
\end{equation}
where $a,b,c,$ and $d$ are some complex parameters and $1\ll\tau_{1}\ll\tau_{2}$.
To be more specific, it is assumed that $\tau_{1}=1/\varepsilon$
and $\tau_{2}=\kappa/\varepsilon^{2}$ with a small parameter $\varepsilon$.
The stability analysis of the linearized system leads to the following
critical pseudo-continuous spectrum 
\[
\lambda=\lambda(\omega,\phi)=\varepsilon^{2}\gamma_{2}(\omega,\phi)+i\omega,
\]
where its real part is determined by the two-dimensional function
(surface) 
\begin{equation}
\gamma_{2}(\omega,\phi)=\frac{1}{\kappa}\left[\ln|c|-\ln\left|i\omega-a-be^{i\phi}\right|\right]\label{eq:2dspectrum}
\end{equation}
(see Sec.~\ref{sub:Spectrum-multiple-delays}). In addition, the
following conditions should be satisfied: $\mbox{Re}\,a<0$ and $|b|<\mbox{Re\,|}a|$;
otherwise, the system is unstable with some stronger instability,
i.e. with positive real part of the spectrum scaling as $\mathcal{O}(\varepsilon)$
or $\mathcal{O}(1)$. A straightforward analysis of the function (\ref{fig:2dspectrum})
shows that it becomes positive when $\mbox{Re}\,a+|b|+|c|>0$. Hence,
one can introduce a natural destabilization parameter $P=\mbox{Re}\,a+|b|+|c|$
with $P=0$ determining the critical value. 

In order to study the dynamics of the system close to this destabilization,
the multiscale anazatz is used leading to the following relation \cite{Yanchuk2014,Yanchuk2015c}
\[
z(t)\approx\varepsilon A\left(\varepsilon^{4}t,\varepsilon t-\nu\varepsilon^{3}t,\varepsilon^{2}t-\nu|b|\varepsilon^{3}t\right),
\]
where $\nu:=-\left(a+|b|\right)^{-1}>0$, the complex amplitude $A(\theta,x,y)$
depends on the slow pseudo-time $\theta=\varepsilon^{4}t$ and on
the two pseudo-spatial variables $x=\varepsilon t(1-\nu\varepsilon^{2})$
and $y=\varepsilon^{2}t(1-\nu|b|\varepsilon)$, each of them corresponding
to the particular delay. In this way, $x$ can be related to $\tau_{1}=1/\varepsilon$
timescale and $y$ to the timescale $\tau_{2}=\kappa/\varepsilon^{2}$
with some additional drift corrections given by the terms of $\varepsilon^{3}$
order. The amplitude $A$ satisfies the normal form equation 
\begin{equation}
\nu^{-1}\Phi_{\theta}=p\Phi+\nu|b|\Phi_{x}-\left(1-\nu|b|^{2}\right)\Phi_{y}+\frac{\nu}{2}\left(\Phi_{xx}+2|b|\Phi_{xy}+|ab|\Phi_{yy}\right)+d\Phi|\Phi|^{2}\label{eq:NF2d}
\end{equation}
with the following boundary conditions 
\begin{equation}
\Phi(x,y,\theta)=e^{i\phi_{b}}\Phi\left(x-1,y,\theta\right),\label{eq:BC1phi}
\end{equation}
\begin{equation}
\Phi(x,y,\theta)=e^{i\left(\phi_{c}-\phi_{b}\left\lfloor \frac{\kappa}{\varepsilon}\right\rfloor \right)}\Phi\left(x-(\frac{\kappa}{\varepsilon}\mbox{mod}1),y-\kappa,\theta\right),\label{eq:BC2phi}
\end{equation}
where $p=P/\varepsilon^{2}$ and $\left\lfloor \cdot\right\rfloor $
denotes the integer part of a number.

If the boundary conditions can be neglected, one can simplify Eq.
(\ref{eq:NF2d}) and eliminate the convective terms with $\Phi_{x}$,
$\Phi_{y}$ as well as the cross-derivative $\Phi_{xy}$, using an
appropriate coordinate transformation. The final equation has the
simpler form: 
\begin{equation}
\Phi_{\theta}=p\Phi+\left|a\right|^{-1}\left(\Phi_{xx}+\Phi_{yy}\right)+d\Phi|\Phi|^{2},\label{eq:GL1}
\end{equation}
with a real diffusion coefficient $\left|a\right|^{-1}$. The simpler
complex Ginzburg-Landau equation (\ref{eq:GL1}) is able to catch
the main qualitative features of the system behavior. The dynamics
of (\ref{eq:GL1}) is known \cite{Cross1993,Chate1996,Aranson2002}
to possess various phase transitions, spiral defects (e.g. for $d=-0.75+i$),
and defect turbulence (e.g. for $d=-0.1+i$); these phenomena can
be observed also in the delay equation for the variable $z(t)$. This
will be discussed in more detail in Sec.~\ref{sub:Spiral-defects-and}.

\subsection{Open questions for normal forms}

The above destabilization theory for long-delay systems takes in account
two basic points: The description of the spectrum determining the
main (co-dimension $1$) destabilizations, such as uniform, wave,
and modulational. The second, the derivation of normal forms which
describe the dynamics within the class of systems, that are undergoing
a particular destabilization transition. 

However, there are still open questions, such as
\begin{itemize}
\item The uniform destabilization is treated for a general class of \emph{scalar}
DDSs (Sec.~\ref{sub:Normal-uniform}), thus the vector case represents
an open problem in general. The case of wave destabilization has been
considered so far for two paradigmatic examples (1D and 2D case) only,
see Secs.~\ref{sub:Turing-instability-1D} and \ref{sub:normalform-2D}.
The normal forms for the general vector case DDS $x'(t)=f(x(t),x(t-\tau))$
are still to be derived. 
\item Determination of the normal form for the modulational instability. 
\end{itemize}

\subsection{Interpretation of delay-equation as a hyperbolic PDE \uline{\label{sub:hyperbolicPDE}}}

The delay-equation (\ref{general}) can be equivalently interpreted
as a hyperbolic PDE by introducing the function of two variables $u(\theta,t)$
as

\begin{equation}
u(\theta,t)=x_{t}(\theta)=x(t+\theta).\label{eq:defu}
\end{equation}
Equation (\ref{general}) can then be written as the following transport
system, where the nontrivial dynamics is introduced mainly by the
boundary condition 
\begin{equation}
\begin{cases}
\frac{\partial u}{\partial t}(\theta,t)-\frac{\partial u}{\partial\theta}(\theta,t)=0, & \theta\in[-\tau,0),\,t\ge0\\
\frac{\partial u}{\partial\theta}(0,t)=F(u(0,t),u(-\tau,t)), & t\ge0\,\,\,\,\,\,\,\,\,\mbox{(boundary cond.)}\\
u(\theta,0)=\varphi(\theta), & \theta\in[-\tau,0]\,\,\,\,\,\mbox{(initial cond.)}.
\end{cases}\label{eq:hyperbpde}
\end{equation}
The first equation follows from the definition (\ref{eq:defu}), the
second one from (\ref{general}) by noting that $x(t)=u(0,t)$, and
the third one is the initial condition. Such a procedure, in contrast
to the reduction to the parabolic PDE discussed in the previous sections
\ref{sub:Reduction-to-parabolic}\textendash \ref{sub:normalform-2D},
does not simplify the problem, and just provides an equivalent representation.
From the point of view of the PDE theory, Eq. (\ref{eq:hyperbpde})
is a hyperbolic initial boundary-value PDE problem, which is especially
challenging since the boundary condition contains a nonlinearity and
a derivative.

\section{The physical meaning of the long-delay limit \label{sec:phys-mean}}

Long-delay systems manifest most of their peculiar features already
in the phenomenology shown by the time series. Numerical simulations
of complex models as well as data from complicated experimental setups
can quickly reveal, when re-organized and analyzed with a suitable
representation the basics elements for the presence of an equivalent
space-time dynamics. While a quantitative description can only be
drawn in terms of specific indicators and in the asymptotic regimes,
preliminary indications often requires only a careful inspection of
the reorganized patterns. 

In this section, we discuss the physical meaning of the indicators
for the space-time dynamics in long-delay systems.

\subsection{Information propagation and causality \label{sub:info-prop}}

\begin{figure}
\begin{centering}
\includegraphics[width=0.5\columnwidth]{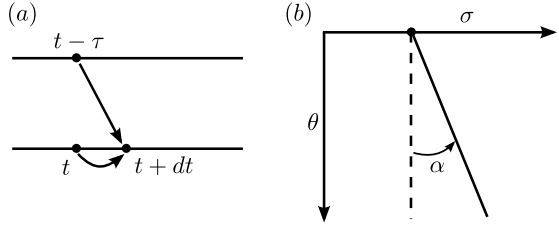} 
\par\end{centering}

\caption{\label{fig:STR-coupling}(a) STR coupling; (b) Comoving angle for
the maximal comoving Lyapunov exponent.}
\end{figure}

As discussed in Sec.~\ref{sec:long delay systems and st dyn}, the
spatio-temporal representation is, in fact, a re-organization of the
temporal series. As long as the spatial size chosen is equal to the
delay time (or sometimes multiple delay times), the propagation of
structures in the pattern is clearly characterized by a \textit{drift},
as illustrated in Fig.~\ref{fig:STR-example}. While it can be quantified
by the analysis of the autocorrelation function and derived from more
fundamental, microscopic quantities, its existence can be understood
using the following simple considerations. In a DDS, the time evolution
is governed by a \textit{local} (i.e., at time $t$) derivative expressed
in terms of the \textit{local} and \textit{past} (i.e., at time $t-\tau$)
terms. When represented in the STR (see Fig.~\ref{fig:STR-coupling}),
it is apparent that the past term cannot influence the dynamics before
(or even at) the time $t$. In other words, there should exist a space-time
bound for the world lines originating in the past (events cone).

The previous considerations can be precised by the analysis of the
propagation of the space-time disturbances, as described by the \textit{maximal
comoving Lyapunov exponent} (MCLE) \cite{Deissler1987,Lepri1996,Giacomelli1996,Lepri1997}.
Generally speaking, the MCLE can be defined as the rate of divergence
(or convergence) of small \textit{localized} perturbations of trajectories
in an extended system, as a function of a suitable comoving angle
$\alpha$ (see Fig.~\ref{fig:STR-coupling}(b)). At variance with
the standard Lyapunov analysis, which is effective for absolute instabilities
only, the MCLE can be used for studying convective instabilities \cite{Briggs1964}
as well. 

In DDS, it has been shown \cite{Giacomelli1996} that using the STR
the MCLE can be defined as in spatially extended systems and its analytical
expression can be derived for the linear model (\ref{eq:scalar}),
obtaining

\begin{equation}
\varLambda(\alpha)=a\sin\alpha+(1+\ln(|b|\tan\alpha)\cos\alpha.\label{eq:MCLE-1}
\end{equation}

\begin{figure}
\begin{centering}
\includegraphics[width=0.8\columnwidth]{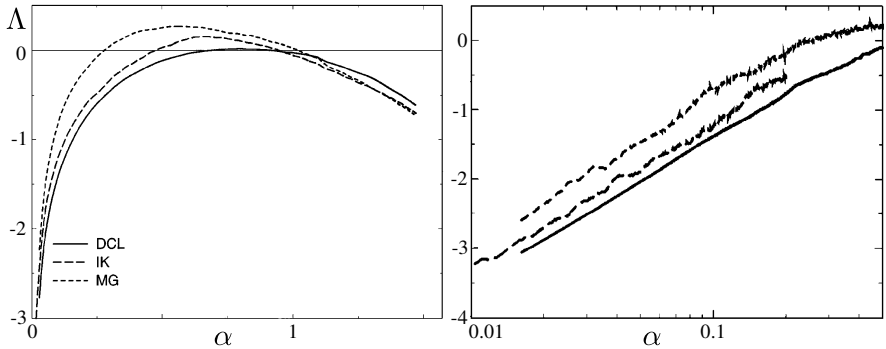} 
\par\end{centering}

\caption{\label{fig:Giacomelli1996} Comoving Lyapunov spectra for the delayed
complex Landau model, Ikeda model, and Mackey-Glass equation (left).
Logarithmic tails are reported in the right figure. (The figure is
adapted with permission from Ref. \cite{Giacomelli1996}. Copyrighted
by the American Physical Society.) }
\end{figure}

From the above expression, we notice the following features:
\begin{itemize}
\item The MCLE is a function of the comoving angle $\alpha$, representing
the inclination of \textit{world}-lines starting at a fixed point
$\sigma$ within the pseudo-space (origin of the disturbance) and
increasing along the pseudo-time $\theta$ (Fig.~\ref{fig:STR-coupling}).
Such a function displays a maximum at a fixed angle $\alpha_{0}$
corresponding to the above-mentioned \textit{drift}.
\item In the case $\alpha\to\pi/2$, the MCLE approaches the coefficient
of the local term. In such a case, it is negative for an effective
STR but it may become positive as well, thus corresponding to the
\textit{anomalous (strongly unstable)} Lyapunov exponent \cite{Lepri1993}
and to the settling of the \textit{strong} chaos regime \cite{Heiligenthal2011}.
\item The MCLE is not defined for $\alpha<0$ and it asymptotically diverges
when $\alpha\to0^{+}$, indicating the bound of the events cone for
the information propagation. The tail at $\alpha=0^{+}$ displays
a logarithmic behavior (see Fig.~\ref{fig:Giacomelli1996}).
\end{itemize}
The bound at $\alpha=0$ is the expression of \textit{causality}:
the system at a given (pseudo) space position can receive information
from the (pseudo) past only from the leftmost spatial positions. Otherwise,
a propagation faster than the evolution speed would be required, i.e.
an ability to travel one delay unit in a time shorter than $\tau$.

The dependence of the MCLE on the propagation angle is similar for
different models, as numerically shown in Fig.~\ref{fig:Giacomelli1996}.
Notably the asymptotics at the causality bound is always logarithmic;
this is related to the finite response time of the system, which is
unable to respond instantaneously at a delayed stimulus. 

\begin{figure}
\centering{}\includegraphics[width=0.4\columnwidth]{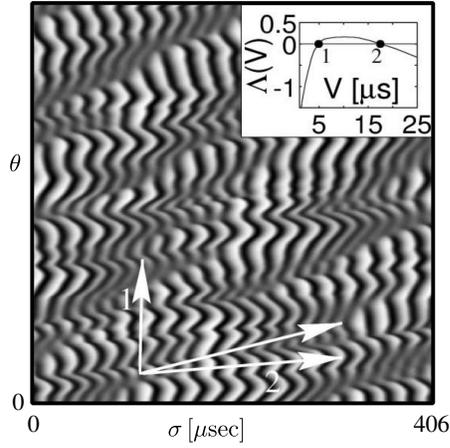}\caption{\label{fig:Giacomelli2000} Space-time representation of the field
intensity in the $CO_{2}$ laser experiment. The horizontal space-like
variable denotes the position within one delay window, while the vertical
time-like variable labels the windows. The central arrow denotes the
direction characterized by the maximal growth rate. Inset: the maximal
comoving Lyapunov exponent. (The figure is adapted with permission
from Ref.~\cite{Giacomelli2000}. Copyrighted by the American Physical
Society.) }
\end{figure}

In general and in presence of \textit{weak} chaos (see Sec.~\ref{sub:Strong-and-weak})
it is expected that the main features of the MCLE remain the same.
An example is presented in Fig.~\ref{fig:Giacomelli2000}, where
the MCLE is calculated from an experimental time series obtained in
a a $CO_{2}$ laser setup \cite{Arecchi1992} using a model-reconstruction
technique especially developed for delay systems \cite{Hegger1998}.
In this example, the horizontal and vertical arrows superimposed to
the STR of the data indicate the directions corresponding to the zeros
of the MCLE (plotted in the inset). The comoving function is evaluated
by the linearized maps obtained from the reconstruction method, as
a direct result from the experimental data. In the paper, it is argued
that the MCLE zeros correspond to macroscopically observable propagation
of structures, since the disturbances neither diverges nor shrink
exponentially along such directions. As a consequence, they should
correspond to the leading contributions to the bidimensional autocorrelation
function of the data. While this is actually the case in examples
of spatially extended systems, in the delayed case only one (the smaller)
of the two velocities corresponds to that found from the correlation
analysis. 

\begin{figure}
\centering{}\includegraphics[width=1\columnwidth]{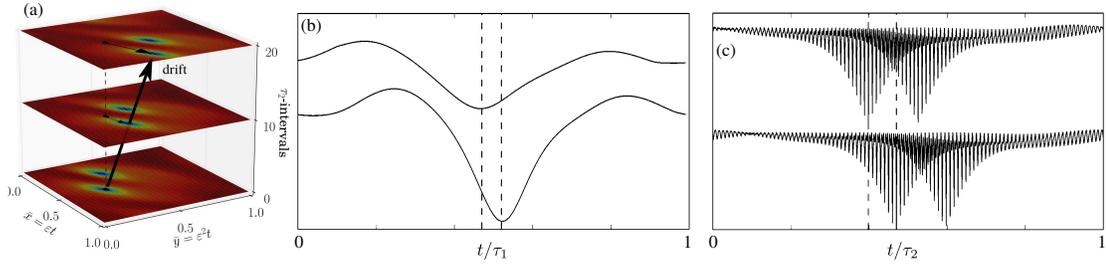} \caption{\label{fig:Yanchuk2015c} (color online) Drift in the propagation
of defect structures in a two-delays system. Using the STR with the
uncorrected pseudo-space variables the defects are moving in the pseudo-time
(a). The components of the (vectorial) drift can be evidenced by plotting
two successive, 2-long (b) and 1-long slices (c). The separations
between the two vertical dashed lines are $y$ (b) and $x$ (c) drift
components respectively. (The figure is adapted with permission from
Ref.~\cite{Yanchuk2015c}. Copyrighted by the American Physical Society.) }
\end{figure}

A generalization to the case of multiple, hierarchically long delays
has been recently presented and discussed in \cite{Yanchuk2015a}.
In the case of two delays, the propagation of structures is illustrated
showing the (vectorial) drift in a two-dimensional STR (see Fig.~\ref{fig:Yanchuk2015c}).
Introducing a spherical coordinates system $(\rho,\alpha,\beta)$
such that the variables of the STR 

\[
t=\sigma+n\tau_{1}+m\tau_{2}
\]
can be written as $m=\rho\cos\alpha,$ $n=\rho\sin\alpha\cos\beta,$
$\sigma=\rho\sin\alpha\sin\beta$, it is found that the MCLE of the
linear system

\[
\dot{z}(t)=az(t)+bz(t-\tau_{1})+cz(t-\tau_{2})
\]
can be calculated explicitly, yielding

\begin{eqnarray}
\varLambda(\alpha,\beta) & = & a\sin\alpha\sin\beta+(1+\ln(|b|\tan\beta))\sin\alpha\cos\beta+\label{eq:MCLE-2}\\
 &  & +(1+\ln(|c|\sin\beta\tan\alpha)\cos\alpha.\nonumber 
\end{eqnarray}
If is apparent that expression (\ref{eq:MCLE-2}) is a two-dimensional
generalization of (\ref{eq:MCLE-1}).

The arrow in Fig.~\ref{fig:Yanchuk2015c} follows the direction defined
by the maximum of $\varLambda(\alpha,\beta)$. We remark that, since
for an arbitrary parameters choice the angles are generically nonzero
and bounded below by $\pi/2$, the disturbances always propagate with
a drift. We notice how the comoving exponent diverges logarithmically
close to the axis $\alpha=0$ and $\beta=0$, i.e. instantaneous propagations
are forbidden. In the opposite limit, $\alpha\rightarrow\pi/2$ (resp.
$\beta\rightarrow\pi/2$), it approaches the value for the single
delay case $c=0$ ($b=0).$ Finally when both $\alpha,\beta\rightarrow\pi/2$
(infinite velocity), $\varLambda=a$ and the dynamics is governed
by the local term as expected.

\subsection{Multiple timescales and the continuum approximation \label{sub:continuum}}

The STR is based on the splitting of the time $t$ in two variables,
the continuous, space-like $\sigma$ within a delay and the integer,
time-like $\theta$ indexing successive delay units; they correspond
to well-separated time scales. In such a representation, the dynamics
if often similar to that of 1-D, spatially extended system as discussed
above: the pattern obtained may show correlations, both in space and
time on possibly long ranges, resulting in a clear and well defined
two-dimensional structure. 

The general context for such a picture, as discussed in Sec.~\ref{sec:math}
is the multiple timescales analysis. Whenever a smooth two-dimensional
pattern can be obtained from the representation, there should exist
a mapping to a spatially-extended model generating the same pattern.
More precisely, the delay system generates a \textit{sampled} sub-pattern,
with integers in the time-like variable, embedded in the pattern generated
by the equivalent spatial model.

The existence of such a mapping is at the basis of the \textit{continuum
approximation}, i.e. the delay pattern appears smooth: the difference
between successive, corresponding spatial points is small, as the
scale for the relevant variations is much larger in the time-like
variable. Conversely, in the case of strong chaos such approximation
fails, as the local instability along the space-like variable (anomalous
or strongly unstable Lyapunov exponent) strongly decorrelates successive
elements in the STR. It should be noted, however, that the strong
chaos is defined in a statistical sense: even if asymptotically the
anomalous exponent exists, it could be possible to build a meaningful
STR for short times, where dynamical fluctuations might set a local,
weak-chaos-like regime.

\subsection{The form of boundary conditions \label{sub:boundary}}

The formal approach for the definition and use of the multiple timescales
including boundary conditions has been discussed in Sec.~\ref{sec:math}.
Here we discuss the form of the bounday conditions in view of the
representation described by Eq.~(\ref{STR}). It holds 
\[
t=\sigma+\theta\tau=(\sigma+\tau)+(\theta-1)\tau
\]
leading to the connection between the values of the solution that
are separated by $\tau$ along the pseudo-spatial variable and by
1 along the pseudo-temporal one:

\[
y(\sigma+\tau,\theta-1)=y(\sigma,\theta).
\]
 By assuming that the solution is slowly changing along the pseudo-temporal
variable $y(\sigma+\tau,\theta-1)\approx y(\sigma+\tau,\theta)$ (e.g.
by strong correlation), and rescaling the spatial variable by $\tau$,
we eventually find the periodic boundary conditions:

\begin{equation}
y(x,\theta)\approx y(0,\theta),\label{SES-bound}
\end{equation}
 besides the appearance of a drift as noted before.

In the case of multiple delays, treated in Sec.\ref{sec:math}, boundary
conditions are more complicated reflecting the interplay of the different
timescales; however, the physical background and the way the systems
approaches the thermodynamic limit is very much the same.

\subsection{Multiple delays and spatio-temporal representation \label{sub:multiple-delays and spatio-temporal representation}}

The spatio-temporal representation can be used as well when the system
contains an arbitrary number of delays, in the case their magnitudes
are ordered hierarchically \cite{Yanchuk2015c}. Consider a dynamical
system with a natural timescale $t_{0}$, with $N$ feedback loops
each with a time lag $\tau_{k}$ $(k=1,...,N)$. We assume that, introducing
the smallness parameter $\epsilon=t_{0}/\tau_{1}\ll1$ the delays
can be written as $\tau_{k}=t_{0}\epsilon^{-k}$. A natural choice
for the multiple scales is now $T_{l}=\epsilon^{l}t$, $l$ being
a positive integer number.

In this case, we expect that $\{T_{l},\,l=1,..,N\}$ represent the
\textit{''spatial}'' scales. The $T_{N+1}$ is now the scale of the
''\textit{drift}'', as measured e.g. microscopically with the help
of the comoving Lyapunov exponent or with the autocorrelation function
\cite{Giacomelli2000}. The expected scale for the equivalent normal
form dynamics is $T_{N+2}$.

In general case, we can define the STR (Eq.~\ref{STR}), with the
variables $\sigma_{0}$, $n_{j}$, and $\Theta$ defined by 
\begin{align*}
[t/\tau_{N}] & =\Theta\\{}
[(t-\Theta\tau_{N})/\tau_{N-1}] & =n_{N-1},\\{}
[(t-\Theta\tau_{N}-n_{N-1}\tau_{N-1})/\tau_{N-2}] & =n_{N-2},\\
...\\{}
[(t-\Theta\tau_{N}-n_{N-1}\tau_{N-1}-...-n_{2}\tau_{2})/\tau_{1}] & =n_{1},\\
t-\Theta\tau_{N}-n_{N-1}\tau_{N-1}-...-n_{1}\tau_{1} & =\sigma_{0}.
\end{align*}
It holds that $\sigma_{0}\in[0,\tau_{1}]$. Because the pseudo-spatial
variables $n_{k}$ can assume large values and are bounded by $\left[\tau_{k+1}/\tau_{k}\right]$,
we use the \textit{\emph{rescaled}}\textit{ pseudo-spatial} variables
$S_{0}=\sigma_{0}/(\tau_{1}/t_{0})$ and $S_{k}=n_{k}/(\tau_{k+1}/\tau_{k})$,
$k=1,\dots N-1$, which are restricted to the interval $[0,1]$, and
the \textit{pseudo-temporal} variable $T=\Theta/(t/\tau_{N})$. We
obtain

\begin{align*}
\sigma_{0}/t_{0}=\sigma_{0}/\tau_{1}\cdot\tau_{1}/t_{0} & =S_{0}\varepsilon^{-1},\\
n_{1}\tau_{1}/t_{0}=n_{1}/(\tau_{2}/\tau_{1})\cdot(\tau_{2}/\tau_{1})\cdot(\tau_{1}/t_{0}) & =S_{1}\varepsilon^{-2},\\
...\\
n_{N-1}\tau_{N-1}/t_{0}=n_{N-1}/(\tau_{N}/\tau_{N-1})\cdot...(\tau_{1}/t_{0}) & =S_{N-1}\varepsilon^{-N}\\
\Theta\tau_{N}/t_{0}=\Theta/(t/\tau_{N})\cdot(t/t_{0}) & =T\varepsilon^{-(N+1)},
\end{align*}
The STR can be written now as 
\begin{equation}
\bar{t}=t/t_{0}=S_{0}\varepsilon^{-1}+S_{1}\varepsilon^{-2}+..+S_{N-1}\varepsilon^{-N}+T\varepsilon^{-(N+1)}.
\end{equation}
It is now easy to recognize how the STR can be related to a multiscale
analysis: the dynamics on the timescale $T_{l}=\varepsilon^{l}t$
is visible on the coordinate $l$ only, since the scales $k<l$ are
too fast and the $k>l$ too slow.

The drift (acting on the scale $T_{N+1}$) and the possible effective
normal form dynamics (evolving in the comoving reference frame, with
scales equal or longer than $T_{N+2}$), can be visualized with the
help of the STR in the pseudo-time range $t/\tau_{N}\simeq\varepsilon^{-1}$
and $t/\tau_{N}\simeq\varepsilon^{-2}$ respectively.

The above approach is a formal framework and represents an interpretation
of the STR that can be built in such a case; however, a rigorous derivation
of the normal forms or even the interpretation of the interplay between
the different time scales will have to be discussed case by case.

\section{Spatio-temporal phenomena in the theory of delay systems \label{sec:examples-theory}}

The choice of a suitable representation, as discussed in the previous
sections, allows to map purely temporal behavior to spatio-temporal
phenomena. In this section, numerical\footnote{\emph{A note on numerical methods:} The numerical integration of delay
systems with fixed delays usually employs the standard methods existing
for ODEs such as Runge-Kutta, etc. The algorithms should take care
of of the saved history from the previous (maximal) delay interval
to manage the integration steps. One of such methods is e.g., pydelay
package for python \cite{FlunkertSchoell2009}. In order to resolve
the fast transitions layers, more advanced algorithms can be adopted,
e.g., in \cite{Giacomelli2012,Giacomelli2013} the authors used an
algorithm of recurrent Taylor expansion of the 30th order with a constant
time step. To numerically exploit bifurcation diagrams, the package
DDE-Biftool is widely applied \cite{Engelborghs2002}. The above utility
allows also the determination of the spectrum of eigenvalues and Floquet
multipliers.} and analytical examples of this approach are presented in the context
of different models and regimes, showing its effectiveness in disclosing
the features of the dynamics hidden behind such complex temporal behaviors.

\subsection{Spatio-temporal chaos in 1D}

A first evidence of spatio-temporal chaos induced by long delay feedback
was reported in \cite{Giacomelli1996,Giacomelli1998} for a Stuart-Landau
oscillator with delayed feedback
\begin{equation}
\frac{dz}{dt}(t)=\mu z(t)-(1+i\beta)z(t)|z(t)|^{2}+\eta z(t-\tau).\label{eq:SL-Gianni}
\end{equation}
The spatio-temporal chaos was shown for $\mu=-0.8$, $\eta=1$, and
$\beta=3,$ in the vicinity of the destabilization $\mu\approx-\eta$.
In such a regime, the amplitude equation (see Sec.~\ref{sub:Turing-instability-1D})
of the form 
\[
\eta\partial_{\theta}A=\mu_{1}A+\frac{1}{2\eta}\partial_{xx}A-(1+i\beta)A|A|^{2}
\]
was derived, where $\mu_{1}=(\mu+\eta)\tau^{2}$. Due to the nonzero
imaginary part of the nonlinearity $\beta\ne0$, the obtained normal
form exhibits spatio-temporal chaos, which is qualitatively similar
to that observed in (\ref{eq:SL-Gianni}) for the corresponding parameter
values.

\subsection{Eckhaus instability \label{sub:Eckhaus-instability} }

\begin{figure}
\begin{centering}
\includegraphics[width=1\linewidth]{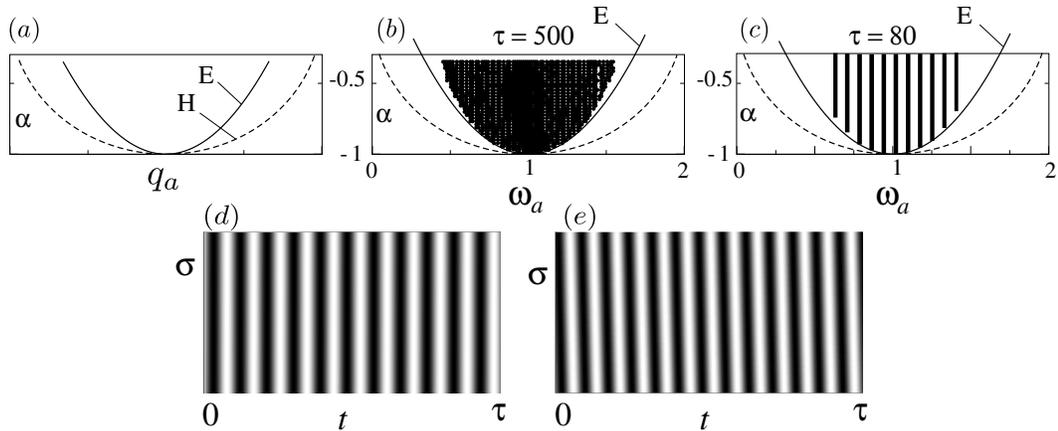}
\par\end{centering}

\caption{\label{fig:Eckhaus}Eckhaus instability in delay system (\ref{eq:Eckhaus-dde}).
(a) Classic Eckhaus instability diagram. $q_{a}$ is the wavenumber
and $\alpha$ is the destabilization parameter. Line $H$ marks where
periodic waves with wavenumber $q_{a}$appear and $E$ is the Eckhaus
line where these patterns become unstable. (b) and (c) are the corresponding
bifurcation diagrams for the delay system (\ref{eq:Eckhaus-dde}),
where the frequency $\omega_{a}$ of periodic solutions plays the
role of the wavenumber. Figures (d) and (e) show the spatio-temporal
representation of the related periodic solutions; the difference is
in the number of peaks (wavenumber). }
\end{figure}

This section summarizes the results presented in \cite{Wolfrum2006},
showing the existence of the Eckhaus destabilization in delay systems.
The Eckhaus phenomenon was first reported in \cite{Eckhaus1965},
where a general framework was proposed for studying the stability
of periodic patterns in Ginzburg-Landau equation, and, in particular
the dependence of their stability on the wavelength. The classic Eckhaus
instability diagram for a PDE on infinite domain is shown in Fig.~\ref{fig:Eckhaus}(a),
where $q_{a}$ is the wavenumber of a periodic pattern ($e^{iq_{a}x}$
for GLE) and $\alpha$ is the destabilization parameter. Increasing
$\alpha$, periodic patterns with wavenumber $q_{a}$ appear at the
line $H$ and become asymptotically stable at the Eckhaus line $E$.
After the destabilization of the homogeneous state, there appear many
stable periodic waves (infinitely for infinite domain, and finitely
many for a bounded domain \cite{Tuckerman1990}). 

In \cite{Wolfrum2006}, it is shown that the delay system 
\begin{equation}
\frac{dz}{dt}(t)=(\alpha+i\beta)z(t)-z(t)|z(t)|^{2}+z(t-\tau)\label{eq:Eckhaus-dde}
\end{equation}
is close to the wave destabilization for $\alpha\approx-1$ (see also
Sec.~\ref{sub:Turing-instability-1D}), and its normal form is the
Ginzburg-Landau equation (\ref{eq:cgl}) where the Eckhaus destabilization
takes place. A corresponding Eckhaus phenomenon is thus present in
DDS (\ref{eq:Eckhaus-dde}), and multiple stable periodic solutions
are found after the destabilization point (for $\alpha$ values larger
than $\alpha=-1$). The number of coexisting stable periodic solutions
is shown to be proportional to the delay $\tau$, hence, the time
delay plays a role similar to the size of the spatial domain (see
Fig.~\ref{fig:Eckhaus}).

We remark that the main difference between the system (\ref{eq:Eckhaus-dde})
and (\ref{eq:SL-Gianni}) is the value of the imaginary part of the
nonlinearity, which is $\beta\ne0$ in (\ref{eq:SL-Gianni}). 

\subsection{Stripes and square waves \uline{\label{sub:Stripes-and-square}}}

\begin{figure}
\centering{}\includegraphics[width=0.8\textwidth]{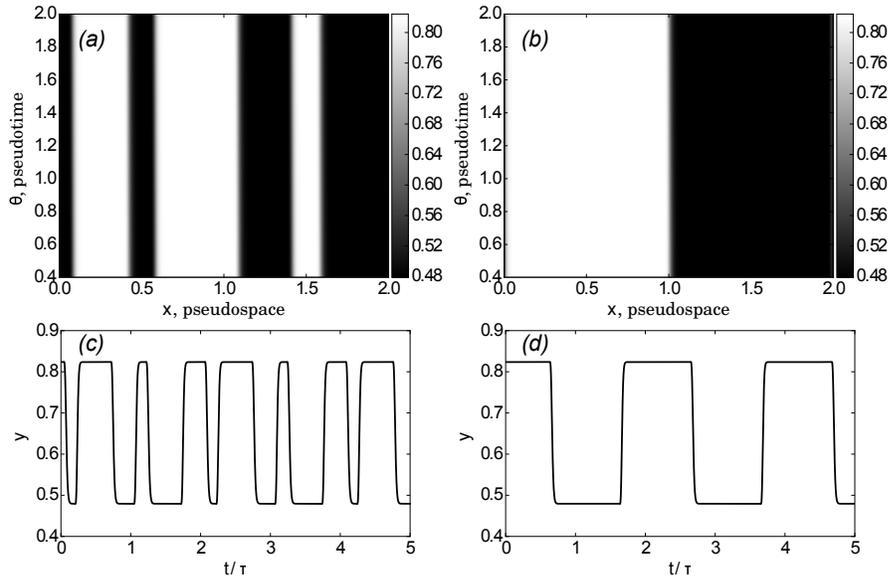}\caption{\label{fig:Square-wave} Square wave solutions (bottom row) of system
(\ref{eq:scalardde})\textendash (\ref{eq:reduced-map}) and their
spatio-temporal representation (upper row). The corresponding reduced
map (\ref{eq:reduced-map}) has a stable period-2 solution that determines
the plateaus of the square-wave solution. Note that the pseudo-spatial
interval of the length $\approx2\tau$ (two consecutive delays) is
used for the spatio-temporal representation, since $y(t+\tau+\delta)\approx-y(\tau)$.
The drift $\delta$ has been taken into account obtaining a vertical
stripes pattern.}
\end{figure}

A remarkable topic in systems with large delays is the existence of
square wave solutions. Their origin has been thoughtfully investigated
theoretically \cite{Mallet-Paret1986,Chow1992,Nizette2004,Nizette2003,Erneux2009,Javaloyes2015}.
In their simplest form, they are shown in Fig.~\ref{fig:Square-wave}(d)
where plateaus of approximately constant lengths close to $\tau$
are connected by sharp fronts. In fact. the period of such a solution
is $2(\tau+\delta)$, where $\delta$ accounts for the drift (see
Sec.~\ref{sec:math}). Higher harmonics could also be observed, where
the plateaus are shorter \cite{Nizette2004}. The bifurcation mechanism
behind the appearance of the multiple coexisting square wave oscillations
resembles the Eckhaus phenomena discussed in Sec.~\ref{sub:Eckhaus-instability}.
Systems, which possess periodic square waves, usually exhibit also
solutions with irregular alternating plateaus {[}Fig.~\ref{fig:Square-wave}(c){]},
which are also ''almost'' $2\tau$ periodic, but in fact they are
very long transients. 

A spatio-temporal representation of such solutions can be made using
the space interval of the length $2(\tau+\delta)$ instead of $\tau+\delta$;
the spatio-temporal patterns associated to the square waves are stripes
as shown in Figs.~\ref{fig:Square-wave}(a) and (b). The square waves
are thus represented by an even set of ascending and descending fronts.
The dynamics of these fronts was to some extent investigated in \cite{Javaloyes2015}
as well as in \cite{Nizette2004}. 

In order to explain the origin of the square wave oscillations, let
us consider the scalar DDS 
\begin{equation}
y'(t)=-y(t)+f(y(t-\tau)).\label{eq:scalardde}
\end{equation}
By rescaling the time $t\to t/\tau$, we obtain 
\[
\frac{1}{\tau}y'(t)=-y(t)+f(y(t-1)).
\]
By formally setting the left-hand side to zero for $1/\tau\to0$,
one obtains the discrete map
\begin{equation}
y(t)=f(y(t-1)),\label{eq:map}
\end{equation}
where, however, $t$ is still the (continuous) time. Now, let us consider
the case when the map (\ref{eq:map}) has a stable period-2 solution
such that $y_{2}=f(y_{1})=f(f(y_{2}))$ with some fixed $y_{1}$ and
$y_{2}$. In such a case, starting from an initial condition, the
map (\ref{eq:map}) will converge to the solution alternating between
$y_{1}$ and $y_{2}$. In particular, the solutions of the form shown
in Figs.~\ref{fig:Square-wave}(c) and (d) appear with exact period
$2\tau$ and infinitely steep fronts between them. As we have seen
from the asymptotic analysis in Sec.~\ref{sec:math}, the neglected
term $\frac{1}{\tau}y'(t)$ adds a diffusion effect and a drift. As
a consequence, among all possible solutions of the map (\ref{eq:map}),
only those survive that have a more smooth profile within the intervals
of almost constant plateaus. Additionally, a drift $\delta$ appears
as well. 

Numerical examples in Fig.~\ref{fig:Square-wave} are presented for
the system (\ref{eq:scalardde}) and for the case

\begin{equation}
f(y)=\mu y(1-y),\label{eq:reduced-map}
\end{equation}
corresponding to the logistic map. The parameter $\mu=3.3$ is chosen
such that the map has a stable period-two solution. 

It is instructive to emphasize the form of the normal form equations
for the DDS (\ref{eq:scalardde}) close to the period-doubling bifurcation
of the reduced map (\ref{eq:map}); there, it holds $f'(y)=-1$, and
the situation is exactly as considered in Sec.~\ref{sub:Normal-uniform}.
Hence, the normal form equation close to such an instability has the
form (\ref{eq:NF}) with the anti-periodic boundary condition $A\left(x,\theta\right)=-A\left(x-1,\theta\right).$
Such a boundary condition corresponds to the flip of the solution
from $y_{1}$ to $y_{2}$ (or vice-versa) after the time interval
$\tau+\delta$. 

The above-mentioned period-doubling bifurcation of the reduced map
leads also to an interesting effect for the spectrum of the steady
state close to the destabilization \cite{Yanchuk2015a}. It corresponds
to the uniform destabilization, see Fig.~\ref{fig:instabilities}
with the leading eigenvalues of the form $\lambda_{k}=i\frac{1}{\tau}\left(\pi+2\pi k\right)+\mathcal{O}(\frac{1}{\tau\text{\texttwosuperior}})$,
$k=0,\pm1,\pm2,\dots$. Such a form of the spectrum explains the emergence
of the $2\tau$-periodic solution at the destabilization of the steady
state as well as the other subharmonic bifurcations. 

Similar situation occurs when the reduced map possesses a periodic
solution with a higher period $n$. In such a case, the spatio-temporal
representation should involve the intervals of a length $n(\tau+\delta)$.

We conclude this section with the following remarks: \\
\textendash{} Stable asymmetric square waves of a period close to
\emph{one} delay are possible as shown in \cite{Weicker2012b}, however,
they are found not in a scalar delay system, but in a system with
two variables, similarly to the system where chimera states are observed,
see Sec.~\ref{sub:Chimera-states}. \\
\textendash{} In \cite{Larger2005}, a comparison of the discrete
reduced map (\ref{eq:map}) with the original delay system (\ref{eq:scalardde})
is discussed and experimentally investigated in an optoelectronic
setup. In particular, a period-three window in the chaotic regime
is observed corresponding to the window for the map.\\
\textendash{} By introducing a second delay, the amplitude of the
square waves can be modulated in a nontrivial way, as discussed in
\cite{Weicker2012a}.

\subsection{Strong and weak chaos \label{sub:Strong-and-weak} }

In Sec.~\ref{sec:LyapunovSpectrum} we have discussed how the Lyapunov
exponents of systems with long time delayed feedback can scale generically
as $1/\tau$ or as $\mathcal{O}(1)$, i.e. do not scale with $\tau$.
In the latter case, if such a Lyapunov exponent is positive, it can
be estimated from the truncated ODE (\ref{eq:trunc}), where the Jacobi
matrix $A(t)$ is evaluated on the solution $s(t)$ of the full delay
system $A(t)=\partial_{1}F(s(t),s(t-\tau))$. The number of such strongly
unstable Lyapunov exponents is limited by the number of components
of $x(t)$. 

When the dynamics has strongly unstable Lyapunov exponents, the regime
can be called \emph{strong chaos}. Indeed, in such a case the perturbations
grow exponentially on time intervals $1/\lambda_{\max}$ that are
much smaller than the delay $\tau$, i.e. the sensitive dependence
on initial conditions reveals on scales much shorter than $\tau$. 

On the other hand, in many situations strongly unstable Lyapunov exponents
are absent. For instance, this happens in the Mackey-Glass \cite{Mackey1977a}
system 
\[
\frac{dx(t)}{dt}=-\gamma x(t)+\beta\frac{x(t-\tau)}{1+x^{n}(t-\tau)},\,\,\gamma>0,
\]
for which $A(t)=-\gamma$ is constant and negative, and, hence, the
truncated ODE $x'(t)=-\gamma x(t)$ cannot have a positive LE. In
such a situation, chaotic attractors possess families of positive
Lyapunov exponents that scale as $1/\tau$. Such a situation is called
\emph{weak chaos}.

\begin{figure}
\begin{centering}
\includegraphics[width=0.8\textwidth]{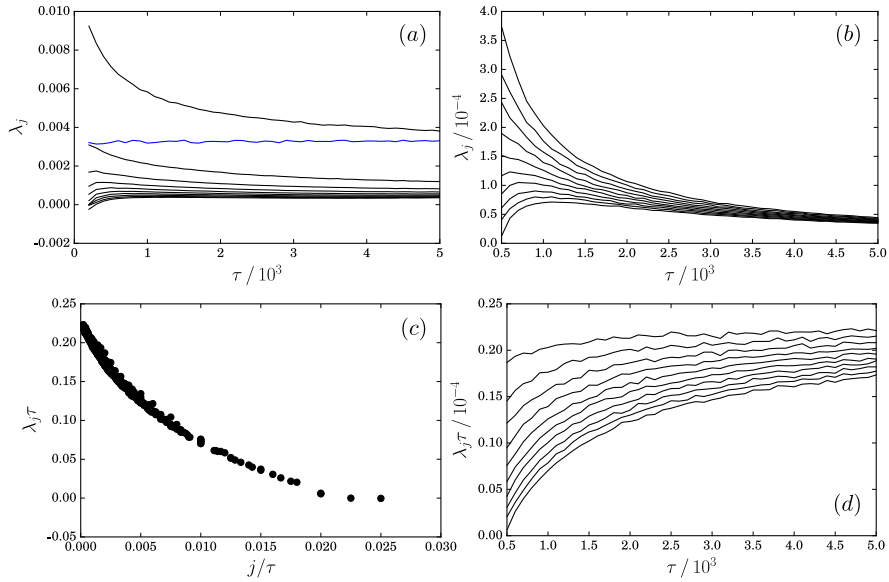}
\par\end{centering}

\caption{\label{fig:LE-strong-weak-chaos}Lyapunov exponents for the Lang-Kobayashi
system (\ref{eq:LK}) versus time delay $\tau$. (a) Strong chaos
for $J=0.075$; (b-d) weak chaos for $J=-0.05$. Other parameters:
$\varepsilon=0.03$, $\eta=0.1$, $\alpha=5.0$, and $\varphi=0.5$. }
\end{figure}

Examples for the Lyapunov exponents in the strong and weak chaos regime
are shown in Fig.~\ref{fig:LE-strong-weak-chaos} for the Lang-Kobayashi
model
\begin{equation}
\begin{array}{c}
E'(t)=(1+i\alpha)N(t)E(t)+\eta E(t-\tau),\\
N'(t)=\varepsilon\left[J-N(t)-(2N(t)+1)|E(t)|\text{\texttwosuperior}\right],
\end{array}\label{eq:LK}
\end{equation}
 where $E(t)$ is the complex field amplitude and the real variable
$N(t)$ the carrier density. This system displays both regimes depending
on the parameters. In particular, by changing the pump current parameter
$J$, the initially stable ''off-state'' $E=0$ destabilizes and,
if the time delay is large enough, increasing $J$ the system quickly
reaches a weakly-chaotic state that is characterized by the family
of positive LEs scaling as $1/\tau$. The dependence of LE's on time
delay is shown in Fig.~\ref{fig:LE-strong-weak-chaos}(b,d). The
LEs approach a continuous spectrum when $\lambda_{j}(\tau)\tau$ is
plotted versus $j/\tau$ where $j$ is the number of LE accordingly
to its magnitude, see Fig.~\ref{fig:LE-strong-weak-chaos}(c). When
the pump current $J$ is further increased, the strong chaos appears,
characterized by a positive maximal Lyapunov exponent that does not
scale with $\tau$, see Fig.~\ref{fig:LE-strong-weak-chaos}(a). 

An useful approach for estimating the scaling of LEs in the transition
between the strong and weak chaos is the stochastic modeling \cite{Dorizzi1987,Juengling2015a}.
In particular, in \cite{Juengling2015a} a linear stochastic model
for the dynamics of a perturbation for a delay system is proposed,
which has a multiplicative noise in its instantaneous part. Such a
model contributes to the explaination of the scaling laws of LEs. 

More details about strong and weak chaos can be found in \cite{Lepri1993,Heiligenthal2011}. 

\subsection{Coarsening \uline{\label{sub:Coarsening}}}

Spatiotemporal dynamics in bistable media is often characterized by
the appearance of fronts that are bounding domains with different
homogeneous states (phases). For example, fronts appear in reaction-diffusion
equations, where local bistable dynamics couples to the spatial degrees
of freedom via a diffusion term. A simplest example is given by a
one-component system \cite{Dee1988}. When such a system is prepared
in an inhomogeneous initial state, the resulting typical patterns
are formed by domains of homogeneous states separated by moving fronts:
they propagate without changing their shape and annihilate in pairs.
This process eventually leads to a globally homogeneous landscape,
a process known as \textit{coarsening} \cite{Bray1994}. Which of
the two phases will eventually fill the whole space, as well as the
propagation speed of the fronts would depend on the shape of the corresponding
double-well potential. 

In \cite{Giacomelli2012}, the coarsening phenomena is shown to take
place in the bistable systems with long time delayed feedback
\begin{equation}
\frac{dx(t)}{dt}=-U'(x(t))+gx(t-\tau),\label{eq:coars}
\end{equation}
where $U(x)$ is a double-well potential with $U'(x)=x(x+1+a)(x-1)$.
For $a=0$ the potential is symmetric and the value of $a$ measures
the asymmetry of $U$. The parameter $g$ determines the strength
of the delayed feedback. In the absence of the feedback, this system
is bistable, i.e. depending on the initial conditions it approaches
one of the two stable steady states. With nonzero feedback, the steady
states are modified as $x_{\pm}=(-a\pm\sqrt{(2+a)^{2}+4g})/2$, and
they remain stable in some range of the parameter $g$. As a result,
for initial conditions $\varphi(s)$ ($-\tau\le s\le0$) that are
close to one of these steady states, the solution will eventually
converge to it. However, when the system is not close initially to
either $x_{-}$ or $x_{+}$, the solution displays oscillations which
rapidly acquire the nearly rectangular form of alternating plateaus
close to the steady states (phases) $x_{\pm}$, separated by ascending
and descending fronts. In the course of time, one of the phases gradually
expands at the cost of the other, until the steady solution gets eventually
established. The ''winning'' solution is determined by the asymmetry
$a$ of the potential. The largeness of the time delay $\tau$ in
this case is necessary for the existence of at least several stationary
phases and the switching episodes between them within one delay interval.

\begin{figure}
\centering{}\includegraphics[width=0.7\textwidth]{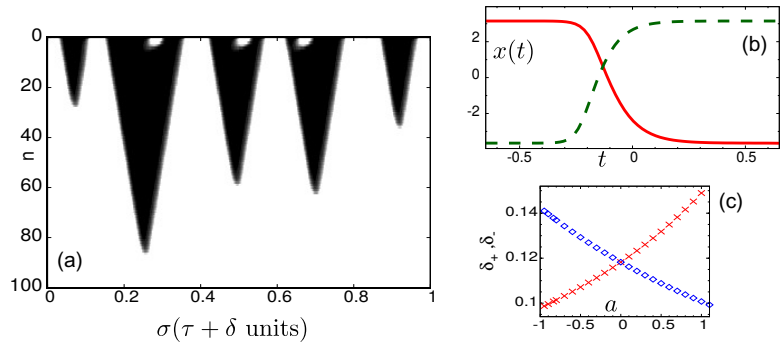} \caption{(Color on-line) (a) Motion of fronts and coarsening in Eq.~(\ref{eq:coars})
at $a=1.5$, $g=1$, and $\tau=10$. (c) Homo- and heteroclinic states
in Eq.~(\ref{eq:coars-profile}) at $a=0.5$, $g=10$. Solid line:
descending front with the velocity $\delta_{-}=0.13183$. Dashed line:
ascending front, $\delta_{+}=0.10882$. (d) Front velocities at $g=10$,
computed from Eq.~(\ref{eq:coars-profile}). Crosses: descending
fronts; diamonds: ascending fronts. \label{fig:coars}}
\end{figure}

A proper interpretation of the phenomenon described above is obtained
using the spatio-temporal representation as described in Sec.~\ref{sub:str}.
A typical spatio-temporal plot for the positive feedback is shown
in Fig.~\ref{fig:coars}(a), where the coarsening of the phases can
be clearly observed. For the given parameter values, the white color
corresponds to the phase with $x_{+}$, and this phase ''wins'' because
of the asymmetry of the double-well potential. The velocities and
the shapes of the fronts are shown in Figs.~\ref{fig:coars}(b-c).
The so called Maxwell point at $a=0$ is also observed, where the
velocities of the ascending and the descending fronts coincide. In
\cite{Giacomelli2012} it is shown that the ''profile equation'' determining
the velocity $\delta$ and the shape of the front is given by the
system with advanced argument 
\begin{equation}
\frac{dy(t)}{dt}=-U'(y(t))+gy(t+\delta).\label{eq:coars-profile}
\end{equation}
The fronts are given as heteroclinic solutions of this equations,
and the velocity $\delta$ is determined from the condition that such
a heteroclinic solution exists. Further analytical results (e.g. scaling
of the coarsening time) are possible to obtain by applying post-newtonian
approximation to (\ref{eq:coars-profile}) assuming that the velocity
$\delta$ is small, see \cite{Giacomelli2012}. 

\subsection{Nucleation}

The term \textit{nucleation} refers usually to a spatial effect in
the context of phase transitions of the first order: the birth of
localized buds of the new phase in the bulk of the old one. Typically,
when the nucleus of the new phase is created inside the old one, the
gain in free energy is proportional to the nucleus volume whereas
the loss is proportional to its surface area. The balance is reached
at a certain critical size of the nucleus, below which the nuclei
shrink \cite{LandauLifshitz}. In the context of reaction-diffusion
systems, nucleation occurs in bistable situations, in which one of
two stable regimes dominates. Here, again, survival and subsequent
growth of the newborn nucleus of the dominating phase require that
at the moment of birth this nucleus occupies a sufficiently large
portion of available space \cite{BaerCh.ZuelickeEiswirthEtAl1992,GomilaColetOppoEtAl2004}.
In one-dimensional, spatially extended systems nucleation has been
studied in e.g. amplitude equations \cite{Aranson2002} or in excitable
media \cite{ArgentinaCoulletMahadevan1997}. 

\begin{figure}[h]
\centering{}\includegraphics[width=0.8\textwidth]{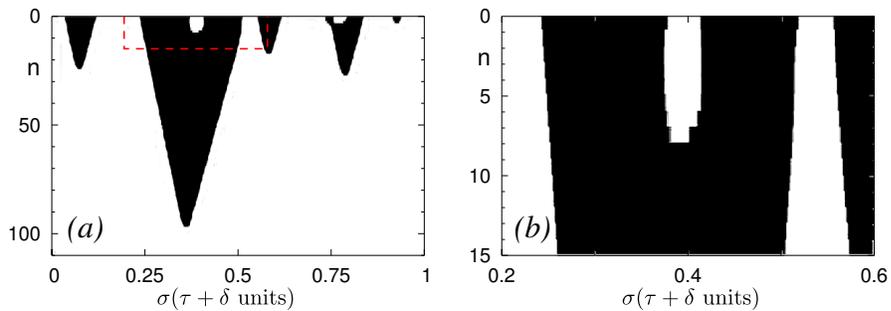} \caption{Coarsening and nucleation in pseudo-space of Eq.~(\ref{eq:coars}).
In the zoom in Fig.~(b), there is a small nucleus of the white phase,
too small to determine the expansion of its phase, even though this
phase is preferred by the choice of the bistable potential (see more
details in \cite{Giacomelli2013}). \label{fig:nucleation} }
\end{figure}

In \cite{Giacomelli2013} the evidence of one-dimensional nucleation
has been given which, in a seeming contrast to the above cases, occurs
not in space but in time. Accordingly, a nucleus occupies not the
spatial region but a time interval, and the critical size is replaced
by a critical duration. This effect takes place in bistable, long-delayed
dynamical systems. In particular, the model as well as the corresponding
spatio-temporal representation is the same as in the previous Sec.~\ref{sub:Coarsening},
and is given by Eq.~(\ref{eq:coars}). Figures~\ref{fig:nucleation}(a)
and (b) illustrate how a small nucleus of the ''winning'' phase shrinks
because of its initial size is not large enough. 

By gradually increasing the width of the nucleus, one can find the
critical nucleus which theoretically exists and is stationary for
an infinitely large time. Dynamically, the critical nucleus can be
viewed as an unstable periodic solution of Eq.(\ref{eq:coars}) with
period $\tau+\delta$ and a single Floquet multiplier outside the
unit circle. The stable manifold of this solution has codimension
1 and serves in the phase space as part of the boundary between the
attraction basins of two fixed points which correspond to the strong
and the weak phases, respectively.

We should remark that for positive values of the delay feedback $g$
the equation (\ref{eq:coars}) is a monotone dynamical system and,
hence, cannot possess stable periodic solutions~\cite{Smith2011}.
However, in the case of vanishing asymmetry $a=0$, the relevant periodic
state can be made nearly neutrally stable, that prolongs noticeably
the life of slightly subcritical nuclei.

\subsection{Pulses and localized structures}

\begin{figure}
\begin{centering}
\includegraphics[width=0.6\textwidth]{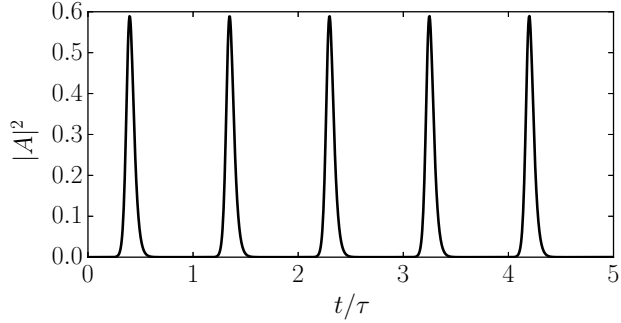}
\par\end{centering}

\caption{\label{fig:MLL}Periodic solution of the model for mode-locked laser
(\ref{eq:MLL}) for parameter values $\gamma=33.3$, $\kappa=0.1$,
$\alpha_{g}=0$, $\alpha_{q}=0$, $\psi=0$, $G_{0}=0.6$, $Q_{0}=1.2$,
$\gamma_{g}=0.0133$, $\gamma_{q}=1$, $s=25$, and $\tau=2.0$. The
solution has the form of narrow pulses with the period close to the
time delay $\tau$. }
\end{figure}

One relevant example of localized solutions existing in the pseudo-space
of systems with time delay is found in delay-models for mode-locked
lasers \cite{Vladimirov2005,Nizette2006}. More specifically in \cite{Vladimirov2005}
the mathematical model for a ring laser with a saturable absorber
is derived in the form of delay-differential equation as

\begin{eqnarray}
\dot{A}(t) & = & \gamma\left(-A(t)+R\left(G(t),Q(t)\right)A\left(t-\tau\right)\right),\nonumber \\
\dot{G}(t) & = & G_{0}-\gamma_{g}G(t)-e^{-Q(t)}\left(e^{G(t)}-1\right)|A\left(t-\tau\right)|^{2},\label{eq:MLL}\\
\dot{Q}(t) & = & Q_{0}-\gamma_{q}Q(t)-s(1-e^{-Q(t)})|A\left(t-\tau\right)|^{2},\nonumber 
\end{eqnarray}
where $R(G,Q)=\sqrt{\kappa}e^{\left(1-\alpha_{g}\right)G/2-\left(1-\alpha_{q}\right)Q/2-i\psi}$,
the variable $A(t)$ is complex and denotes the amplitude of the electric
field, and the real variables $G(t)$ and $Q(t)$ stand for the saturable
gain and loss respectively. For certain parameter values, this system
exhibits localized periodic pulses with the period close to $\tau$,
see Fig.~\ref{fig:MLL}. The emergence of several pulses within the
delay interval in this system is studied in \cite{Nizette2006}. 

Other examples of solutions that are localized within the delay interval,
and hence, localized in the spatio-temporal representation, appeared
recently in the papers \cite{Garbin2015,Marconi2015}. In particular,
in \cite{Garbin2015} the authors report solutions which are essentially
localized phase defects. The theoretical model which reproduces this
effect was shown to be the Ginzburg-Landau equation including an external
injection and feedback terms. In the case when the injection and feedback
are weak, the model was reduced to the delay equation 
\[
\frac{d\varphi(t)}{dt}=\Delta-\sin\varphi(t)+\kappa\sin(\varphi(t-\tau)-\varphi(t)-\psi),
\]
where $\Delta,$ $\kappa$, and $\psi$ are parameters. For an appropriate
choice of these parameters and a large delay $\tau$, the system exhibits
a sharp localized phase jump, which occurs within every delay interval.
By using multiple scale analysis, the authors in \cite{Garbin2015}
derive the amplitude equation in the form of the modified Sine-Gordon
equation 
\[
\frac{\partial A}{\partial\theta}=\sin\bar{\varphi}-\sin\varphi+\frac{\partial A}{\partial x\text{\texttwosuperior}}+\tan\psi\left(\frac{\partial A}{\partial x}\right)^{2},
\]
where $\sin\bar{\varphi}=\Delta-\sin\psi$, $x$ is the pseudo-space
and $\theta$ is the pseudo-time. As a necessary conditions for the
appearance of such a localized solution, the authors mention the combination
of the saddle-node bifurcation on a circle and a delayed feedback.

\subsection{Spiral defects and turbulence in 2D \label{sub:Spiral-defects-and}}

Many new challenging problems arise when a system is subject to several
delayed feedbacks acting on different timescales \cite{Yanchuk2014,Yanchuk2015c}.
In contrast to the single delay situation, essentially new phenomena
occur, related to higher spatial dimensions involved in the dynamics. 

A simple paradigmatic setup for the multiple delays case is the following
system 
\begin{equation}
\dot{z}=az+bz(t-\tau_{1})+cz(t-\tau_{2})+dz|z|^{2},\label{eq:SL2delays}
\end{equation}
which generalizes the single feedback case (\ref{eq:Eckhaus-dde}).
Equation (\ref{eq:SL2delays}) describes the interplay of the oscillatory
instability (Hopf bifurcation) and two delayed feedbacks, which we
consider acting on different timescales $1\ll\tau_{1}\ll\tau_{2}$.
The variable $z(t)$ is complex, and the parameters $a$, $b$, and
$c$ determine the instantaneous, $\tau_{1}$-, and $\tau_{2}$-feedback
rates, respectively. 

In \cite{Yanchuk2014,Yanchuk2015c} it is shown that two-dimensional
patterns such as spiral defects and defect turbulence are typical
of system (\ref{eq:SL2delays}). Moreover, they can be also generically
found in a semiconductor laser model with two optical feedbacks.

The main idea is that the spatio-temporal representation (see Sec.~\ref{sub:str})
with 2D spatial domain can be used since the time delays are acting
on different timescales. The first spatial coordinate is related to
the timescale $\tau_{1}$, and the second one to $\tau_{2}$. The
corresponding normal form is the 2D Ginzburg-Landau equation (\ref{eq:GL1})
with some specific boundary condition (see Sec.~\ref{sub:normalform-2D}).
As a result, the 2D spatiotemporal phenomena known from partial differential
equations \cite{Chate1996,Cross1993,Cross2009} can be found in delay
systems of the form (\ref{eq:SL2delays}). 

\begin{figure}
\includegraphics[width=1\columnwidth]{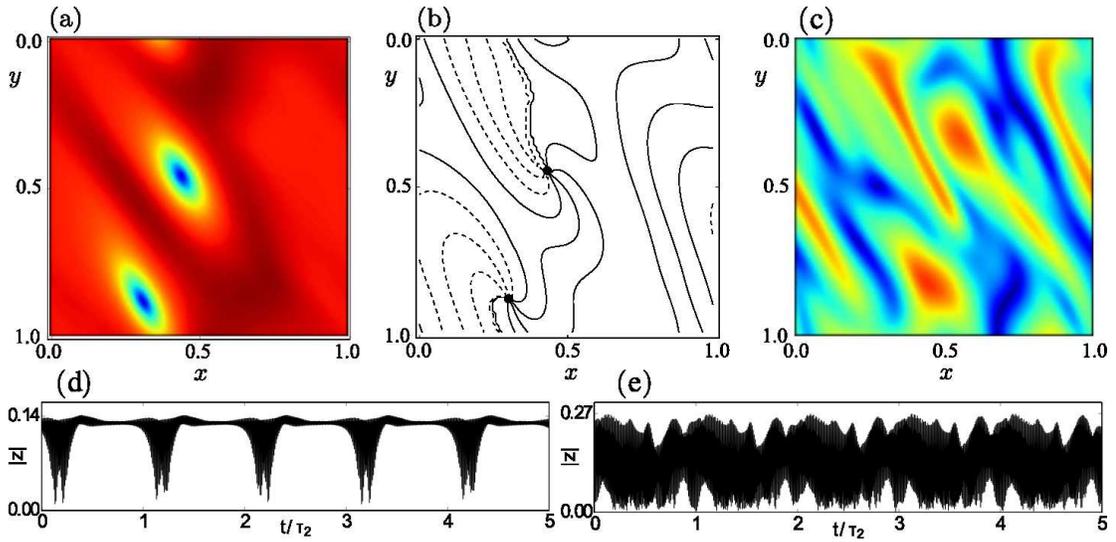}

\caption{\label{fig:2d} Spiral defects in system with two delays (\ref{eq:SL2delays}).
(a-b) 2D spatio-temporal representation of the time series for $|z|$
and the phase of $z$, respectively, reveals the spiral defects. Shown
are snapshots for a fixed value of the pseudo-spatial coordinate $\theta$.
In (a), the magnitude of $|z|$ shown in color, and in (b), the constant
level lines for the phase are plotted. (c) Similar spatio-temporal
representation of the time series for other parameter values (see
text for the parameters) of $|z(t)|$ reveals the defect turbulence.
(d), (e) Time series corresponding to the defects (a-b) and turbulence
(c), respectively.}
\end{figure}

Figure~\ref{fig:2d} shows spiral defects and defect turbulence found
in Eq.~(\ref{eq:SL2delays}). The associated time-series $|z(t)|$
are shown in Figs.~\ref{fig:2d}(d) and \ref{fig:2d}(e), computed
for the parameter values $a=-0.985$, $b=0.4$, $c=0.6$, $d=-0.75+i$,
$\tau_{1}=100$, and $\tau_{2}=10000$ for Fig.~\ref{fig:2d}(a,b,d)
and the same parameters except $d=-0.1+i$ for Fig.~\ref{fig:2d}(e,c).
These parameter values correspond to the situation where the system
is slightly above the instability threshold ( $a_{thr}=-1$), and
the zero steady state is weakly $\tau_{2}$-unstable, i.e. the only
unstable spectrum is the $\tau_{2}$-spectrum (see Sec.~\ref{sub:Spectrum-multiple-delays}). 

The time series exhibit oscillations on different timescales related
approximately to the delay times. However, the two-dimensional spatio-temporal
representation of the data (see Sec.~\ref{sub:str}) in Figs.~\ref{fig:2d}(a-c)
reveal the nature of the dynamical behaviors. In particular, the first
case in \ref{fig:2d}(a-b) corresponds to a (frozen) spiral defects
solution, where the two coexisting spiral defects are shown by the
dots. At the defect points, the level lines for the phase meet, the
phase is not defined, and $|z|=0$. The solution shown in Fig.~\ref{fig:2d}(c)
corresponds instead to the defect turbulence regime, where the modulation
of the amplitude $\left|z(t)\right|$ starts to approach the zero
level in a random-like manner. The corresponding spatial representation
reveals non-regular motions of the spiral defects. The plots are snapshots
in time. 

Those types of defects and turbulence regimes have been also reported
for the Lang-Kobayashi model of a single-mode, semiconductor laser
with two optical feedbacks \cite{Yanchuk2014}.

\section{Experiments on long-delay systems\label{sec:examples-exp}}

In this section we review a series of experiments on long-delayed
feedback systems. Their spatio-temporal nature is disclosed with a
proper representation and a rich variety of phenomena appear, which
a purely temporal description is unable to show. Most of the experiments
described have been performed in optical systems, in particular in
laser setups.

\subsection{Spatio-temporal chaos and defects turbulence}

The effect of a delayed feedback in the case of short delays was investigated
in a semiconductor laser (where the propagation delay is naturally
relevant because of the short timescales of the system) \cite{Giacomelli1989}
and in a $CO_{2}$ laser \cite{Arecchi1991}. In the former case,
a Hopf bifurcation followed by stable oscillations was observed, with
frequencies well reproduced by a linear model; in the latter, a far
more complicated behavior with period doubling, frequency locking
and quasi-periodicity leading to chaos was shown.

\begin{figure}
\centering{}\includegraphics[width=1\columnwidth]{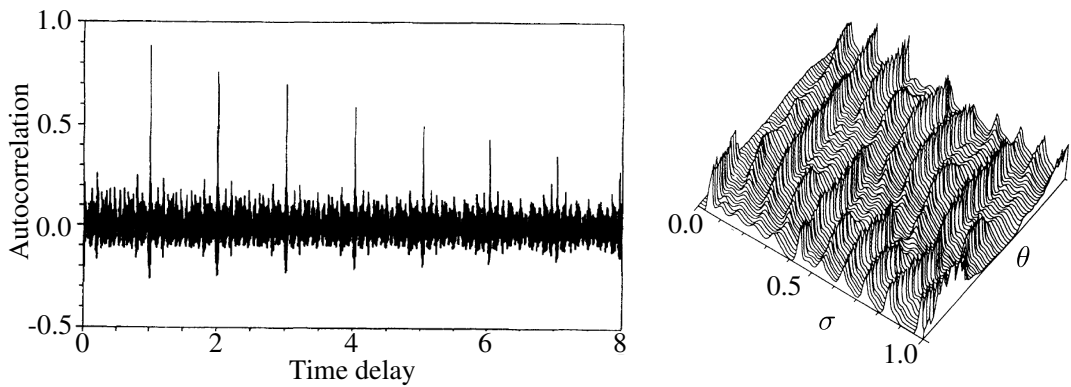}\caption{\label{fig:Arecchi1992} Normalized autocorrelation function of the
$CO_{2}$ laser intensity (left); the horizontal scale is in units
of the delay time. Right: spatio-temporal representation of the signal.
(The figure is adapted with permission from Ref.~\cite{Arecchi1992}.
Copyrighted by the American Physical Society.) }
\end{figure}

Later on, the same group performed a study where the delay time in
the feedback loop on the $CO_{2}$ laser was greatly increased due
to a specific setup \cite{Arecchi1992}. The resulting dynamics showed
again a sequence of complicated features, but the analysis of the
autocorrelation function indicated the existence of at least two well-separated
time scales (see Fig.~\ref{fig:Arecchi1992}, left). A data reorganization,
as pictured in Eq.~(\ref{STR}) allowed a clear understanding of
the role of the different time scales. The bidimensional pattern obtained
(Fig.~\ref{fig:Arecchi1992}, right), displays a complicated space-like
dynamics within a delay cell evolving along the time-like direction. 

In that paper, the STR was first introduced and used to describe the
spatio-temporal properties on an experimental system with long delay.
In such a representation, the features of a transition towards (pseudo)
space-time chaos were shown, as a function of a system's parameter.

\begin{figure}
\begin{centering}
\includegraphics[width=1\columnwidth]{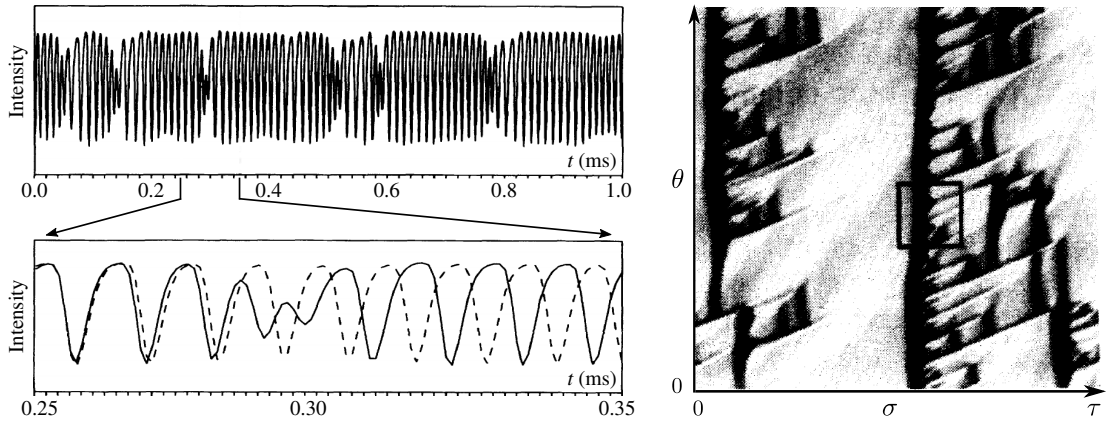} 
\par\end{centering}

\caption{\label{fig:Giacomelli1994} Temporal series of the $CO_{2}$ laser
intensity, showing the existence of phase defects (left). Right: spatio-temporal
representation of the signal; the horizontal scale is one delay time,
the vertical is the pseudo time. (The figure is adapted with permission
from Ref.~\cite{Giacomelli1994}. Copyrighted by the American Physical
Society.) }
\end{figure}

In a successive work \cite{Giacomelli1994}, the evidence and characterization
of defect-mediated turbulence was shown in the same setup, with a
transition to a phase turbulent regime. The time series of the laser
intensity evidenced the existence of phase defects, close to a first
destabilization of the steady state via a Hopf bifurcation (Fig.~\ref{fig:Giacomelli1994},
left). The STR showed the propagation of the defects structures (Fig.~\ref{fig:Giacomelli1994},
right), with an erratic behavior. For a choice of the parameters,
the defects regime was a transient phenomenon, with a scaling of the
transient time (in delay units) very close to that predicted for a
diffusive behavior in 1D extended systems.

\begin{figure}
\begin{centering}
\includegraphics[width=0.75\columnwidth]{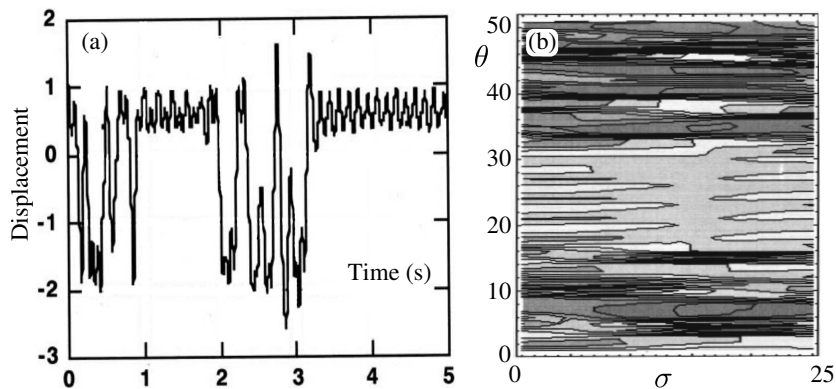} 
\par\end{centering}

\caption{\label{fig:Hikihara1999} (from \cite{Hikihara1999}) Dynamics of
magneto-elastic system with delayed feedback control; (a) displacement,
(b) contour plot of waves on the spatiotemporal state space. (Reprinted
from {[}T.~Hikihara and Y.~Ueda, ''An expansion of system with time
delayed feedback control into spatio-temporal state space'', Chaos
9, 4 (1999), pp. 887-892{]}, with the permission of AIP Publishing.)}
\end{figure}

A widely used technique for the control of chaos was introduced by
Pyragas \cite{Pyragas1992}. The method employs a delayed variables
to stabilize periodic orbits in a chaotic regime. For sufficiently
long delays, the investigation of the effect of such a feedback can
be studied using the STR. In an experiment with a magneto-elastic
beam system under time delayed feedback control \cite{Hikihara1999},
such a method has been applied, plotting the displacement as a variable
(Fig.~\ref{fig:Hikihara1999}(a)) and the corresponding STR (Fig.~\ref{fig:Hikihara1999}(b)).

The autocorrelation function and its structure has been proven to
be a powerful tool to characterize the long-delayed systems; conversely,
the existence of well defined features such that the presence of revivals
and the indication of a drifting is a signature of the possible spatio-temporal
dynamics.

\begin{figure}
\centering{}\includegraphics[width=1\columnwidth]{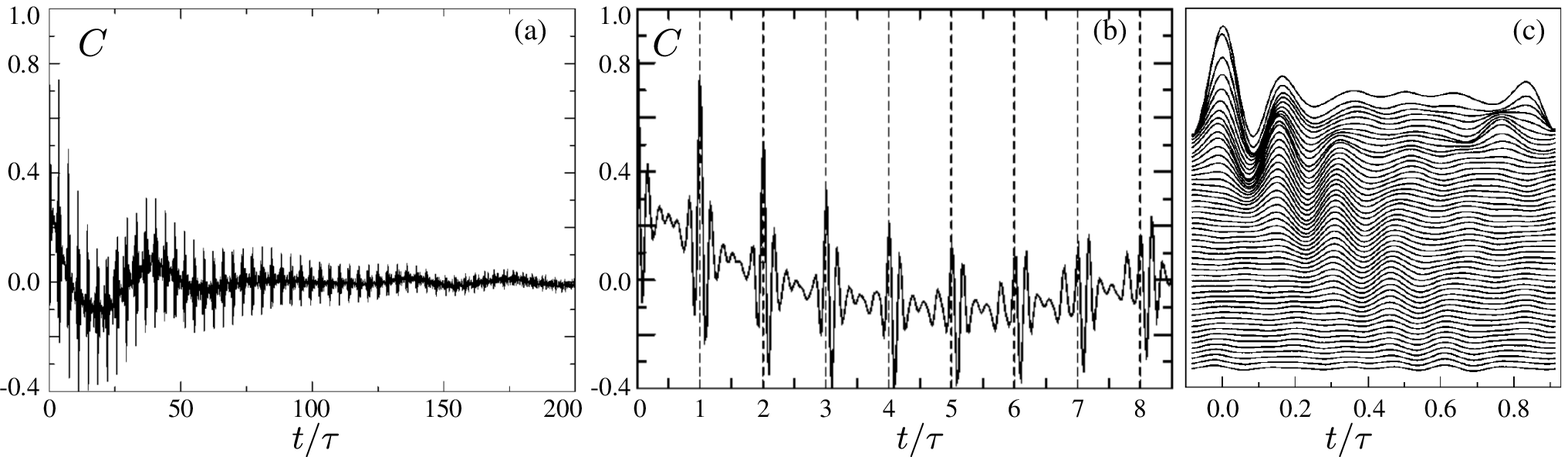} \caption{\label{fig:Giacomelli2003} Autocorrelation function of VCSEL intensity
(a), its zoom (b), and spatio-temporal representation (c). (The figure
is adapted with permission from Ref.~\cite{Giacomelli2003}. Copyrighted
by the American Physical Society.) }
\end{figure}

An example of the use of this indicator is presented in a study of
an experiment carried out in a Vertical Cavity Surface Emitting Laser
(VCSEL) with polarized optical feedback \cite{Giacomelli2003}. The
autocorrelation function of the laser intensity signal displays the
revivals at almost the multiples of the delay time $\tau_{c}$ (Fig.~\ref{fig:Giacomelli2003}
(left) and the inset). In this work, it is suggested to use the STR
for the autocorrelation function (Fig.~\ref{fig:Giacomelli2003},
right): the averaging process of the autocorrelation allows to display
features (in this case, a drift) which are otherwise hidden by the
small timescales and the low signal-to-noise ratio. The delay time
was $\tau_{c}=3.63~ns$, much longer of the scales of spiking dynamics
typical for the optical feedback system which is in the $ps$ range.

\begin{figure}
\centering{}\includegraphics[width=1\columnwidth]{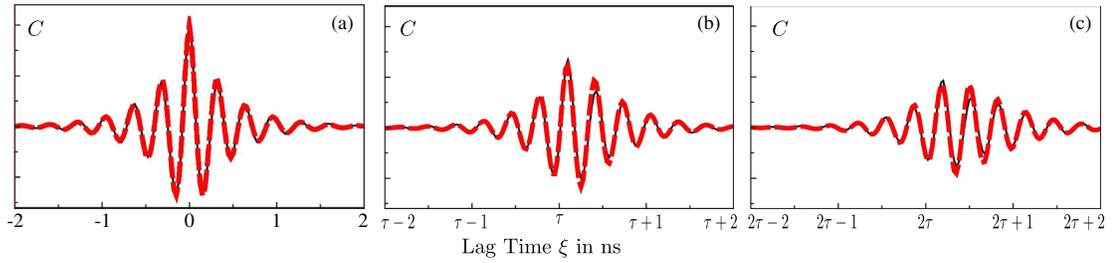} \caption{\label{fig:Porte2015b} (color online) The central peak (a) and the
first two delay echoes {[}(b) and (c), respectively{]} of the autocorrelation
of a semiconductor laser subject to delayed feedback. The solid thin
line shows the experimental data and the dashed lines correspond to
the fitted analytic expressions from a linear stochastic model. (The
figure is adapted with permission from Ref.~\cite{Porte2014b}. Copyrighted
by the American Physical Society.) }
\end{figure}

A further step has been taken in understanding the properties and
using the features of the autocorrelation function in Ref.~\cite{Porte2014b}.
There, the study of a simple linear, stochastic model allowed both
to reproduce the experimental findings from a semiconductor laser
with an optical feedback (see Fig.~\ref{fig:Porte2015b}) and to
evaluate the drift in terms of phenomenological parameters.

\begin{figure}
\centering{}\includegraphics[width=1\columnwidth]{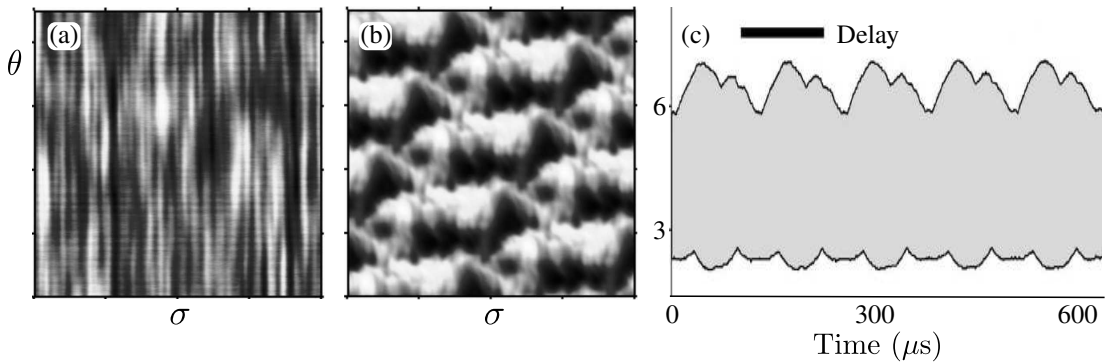}\caption{\label{fig:Franz2008} Spatio-temporal representations of the intensity
dynamics for a laser with (a) no feedback and (b) with the delayed
feedback. (c) Envelope of the time series of the intensity dynamics:
the bar shows the delay $\tau$. (The figure is adapted with permission
from Ref.~\cite{Franz2008}. Copyrighted by the American Physical
Society.) }
\end{figure}

A fiber ring laser, with a feedback loop realized with an additional
fiber ring cavity may represent a convenient, all-optical dynamical
system for studying the effect of long delays. In Ref.~\cite{Franz2008}
this setup produced a large variety of behaviors ranging from the
typical phenomenology expected in short delays, with frequency locking
and low dimensional chaos, to pattern formation and structure propagation
typical for spatio-temporal systems (Fig.~\ref{fig:Franz2008}(b)).
An extension to the case of two long-delayed coupled lasers is also
studied. In Fig.~\ref{fig:Franz2008}(c), example of time series
is shown indicating how the multiple time scales of the system are
folded in the dynamics. In such a system, a suitable decomposition
in terms of the Karhunen-Loevè modes allowed to estimate some of the
statistical indicators of complexity (such as the entropy) and their
scaling with the delay time.

\subsection{Front dynamics: coarsening, nucleation, and pinning}

\begin{figure}
\centering{}\includegraphics[width=1\columnwidth]{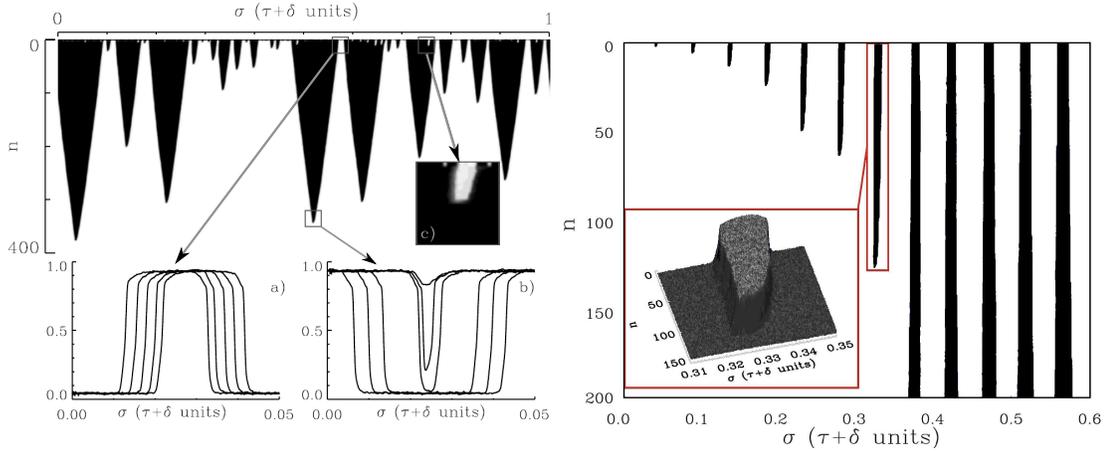} \caption{\label{fig:Giacomelli2012} Left: Spatio-temporal representation of
the polarized laser intensity showing coarsening towards the lower
state (white), with fronts creation (a) and annihilation (b), as well
as nucleation of the strong state into the weak one (c). Right: Spatio-temporal
representation illustrating nucleation for different initial pulse
widths. (The figure is adapted with permission from Ref.~\cite{Giacomelli2013}.
Copyrighted by the American Physical Society.)}
\end{figure}

A hysteresis and a diffusive coupling are the main ingredient for
the so-called bistable reaction-diffusion systems (see e.g. \cite{Nicolis1977}).
A bistable VCSEL with a long-delayed opto-electronic feedback has
shown to present most of the features of the reaction-diffusion systems
\cite{Giacomelli2012}. Depending on the parameters, the formation
and annihilation of pairs of fronts was observed, with coarsening
towards a homogeneous (strong) state (see Fig.~\ref{fig:Giacomelli2012}).
The characterization of the drift in terms of the fronts motions and
its estimation using a simple model has been also provided, as a further
evidence of the typical behavior of long delayed systems. 

The phenomenon of nucleation has been also observed \cite{Giacomelli2012,Giacomelli2013}:
the formation of localized buds of the strong phase into regions of
the weak one (Fig.~\ref{fig:Giacomelli2012}, right panel).

\begin{figure}
\centering{}\includegraphics[width=1\linewidth]{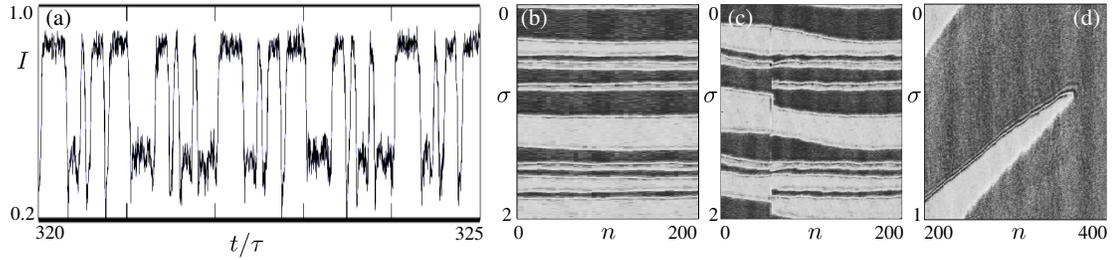}\caption{\label{fig:Javaloyes2015} (a) Time series of the laser intensity
and spatio-temporal representation using a spatial cell of length
$2\tau$ (b,c). Example of domain creation and propagation without
coarsening (c); for other parameters, coarsening occurs (d). (The
figure is adapted with permission from Ref.~\cite{Javaloyes2015}.
Copyrighted by the American Physical Society.) }
\end{figure}

A mechanism for stabilizing temporal domain walls away from the Maxwell
point, based on antiperiodic regimes in a delayed system close to
a bistable situation, leads to a cancellation of the average drift
velocity. As a consequence, the coarsening mechanism is blocked. The
results are demonstrated \cite{Javaloyes2015} in a normal form model
and experimentally in a laser with optical injection and delayed feedback.
In Fig.~\ref{fig:Javaloyes2015}, it is shown how a STR using a pseudo
space cell of different lengths (here $2\tau$is used) allows to recover
the standard information about the propagation of the patterns. In
this case, the coarsening is blocked and the structures propagate
without shrinking or expanding, as seen also in Fig.~\ref{fig:Javaloyes2015}(c)
in the case of the creation of new domains. For different parameter
values, a standard coarsening regime is shown.

\begin{figure}
\centering{}\includegraphics[width=0.75\columnwidth]{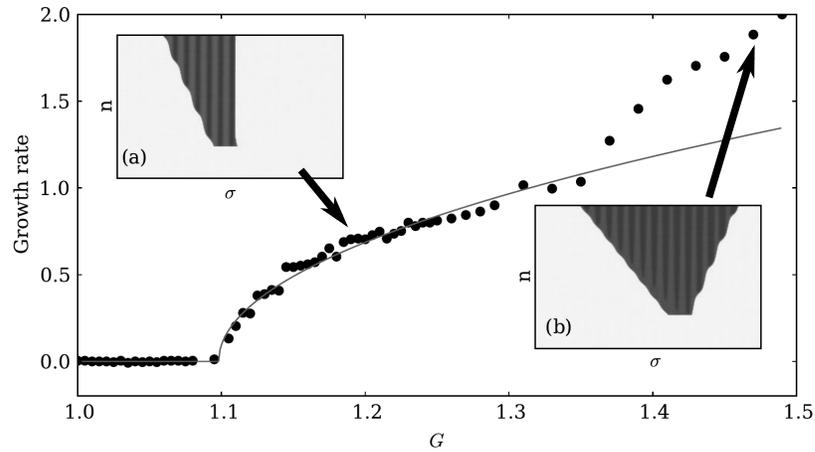} \caption{\label{fig:Marino2014} Pinning regimes, illustrated in terms of the
growth rate of the dominant phase against the bistable states asymmetry
parameter $G$. Solid line: square-root fit in the range $1\le G\le1.3$.
Insets: Space-time representation of the laser intensity for $G=1.2$
{[}(a){]} and $G=1.47$ {[}(b){]}. (The figure is adapted with permission
from Ref.~\cite{Marino2014}. Copyrighted by the American Physical
Society.) }
\end{figure}

The application of a modulated signal produced another typical phenomenon
of extended systems: the pinning/unpinning of localized structures
\cite{Marino2014}. The phenomenon has been shown experimentally in
the setup of \cite{Giacomelli2012}, with the addition of a periodic
modulation on the pump of the VCSEL; a study on a phenomenological
model also confirmed the features of the measurements. In Fig.~\ref{fig:Marino2014},
the experiment evidences the unpinning changing the asymmetry parameter
of the bistable states. For small values, the fronts do not propagate
at all and the growth rate is zero. Above a first threshold, the growth
rate suddenly increases with a square-root scaling, indicating that
the unpinning transition has occurred. However, only the left front
is propagating, whereas the right front is still pinned {[}see inset
(a){]}. The situation remains qualitatively unaltered until a second
bifurcation takes place, illustrated by the abrupt change in the growth
rate. This corresponds to the unpinning of the right front {[}inset
(b){]}. The other front now also drifts away, such that the high power
state invades the whole system.

\subsection{Strong and weak chaos}

The experimental demonstration of the existence of an anomalous (strongly
unstable, see Sec.~\ref{sec:LyapunovSpectrum}) Lyapunov exponent
represents a difficult task. The scaling of the maximal exponent with
the delay time cannot be simply revealed by a visual inspection of
the patterns or even from the evidence of small amplitudes of the
autocorrelation revivals. In fact, the evidence of a transition between
a dynamics with almost coherent propagation in the pseudo time to
a locally strong, chaotic represents only an indication of the possible
existence of the anomalous exponent.

\begin{figure}
\centering{}\includegraphics[width=1\linewidth]{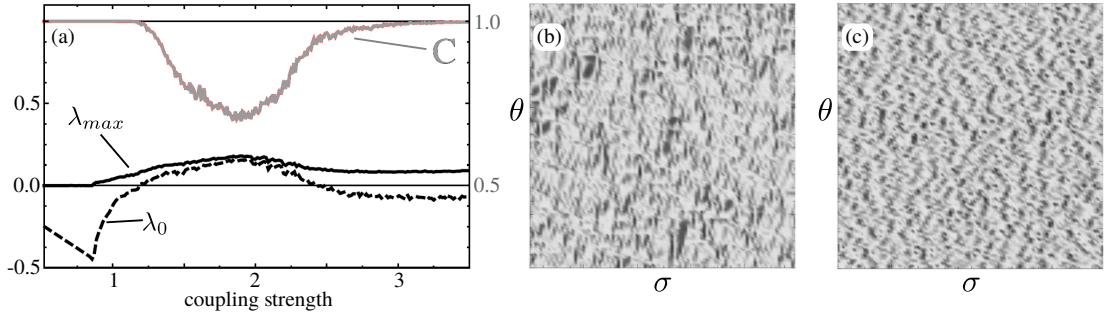}\caption{\label{fig:Heiligenthal2011} (a) Simulated and maximal $\lambda_{\max}$
(solid line) and instantaneous $\lambda_{0}$ (dashed line) Lyapunov
exponents of two electronic circuits and experimentally measured cross-correlation
$C$ {[}gray line{]} between the maxima of the time series versus
coupling strength. (b-c): Space-time diagram of a laser intensity
for a single laser with a self-feedback of $\tau=10ns$. Panel (b)
corresponds to weak chaos and (c) to strong chaos. (The figure is
adapted with permission from Refs.~\cite{Heiligenthal2011} and \cite{Heiligenthal2013}.
Copyrighted by the American Physical Society.) }
\end{figure}

The notion of strong and weak chaos was introduced in \cite{Heiligenthal2011},
discussing the dynamics of complex networks with delayed couplings.
The authors distinguished the two regimes by the scaling properties
of the maximum Lyapunov exponent: the existence of an anomalous exponent
corresponds to the strong chaos regime (see Sec.~\ref{sub:Strong-and-weak}
for more details). It was shown that this is related to the condition
for stable or unstable chaotic synchronization. The concept is illustrated
in simulations of laser models and in experiments with coupled electronic
circuits, and the transitions from weak to strong chaos and vice versa
changing the coupling strength are described. The analysis of the
stability of synchronization between two units shows that chaos is
weak if and only if the two units can be synchronized. Hence, it represents
a less demanding condition for checking the existence of the anomalous
LE. Figure \ref{fig:Heiligenthal2011}(a) shows the simulated LEs
in comparison with the experimentally measured cross-correlation $C$
between the maxima of the time series of the two electronic circuits
as a function of the coupling strength. For small coupling, zero-lag
synchronization of periodic dynamics is observed. If the coupling
is increased, the dynamics becomes chaotic while complete synchronization
is maintained. With a further increase of the parameter, the cross
correlation first decreases and then increases again until synchronization
is reached once more, indicating transitions from weak to strong chaos
and back to weak chaos.

In \cite{Heiligenthal2013} a detailed study of an experiment with
a semiconductor laser with optical feedback and a corresponding model
(Lang-Kobayashi equations \cite{Lang1980}) showed a good indication
that the system could present a transition between weak and strong
chaos (see e.g. Fig.~\ref{fig:Heiligenthal2011}(b-c)). While the
reported results are not definitive, the possibility of a characterization
of the strong chaos regime in a fast, optical system is promising,
even in view of possible statistical analysis given the huge numbers
of samples that can be collected in a single measurement. 

In the same setup, a different approach is presented in \cite{Oliver2015}.
The study focuses on a characterization based on the consistency indicator:
it estimates the reproducibility of responses of a dynamical system
when repeatedly driven by similar inputs, starting from different
initial conditions. The feedback loop is realized with an optical
fiber that can be switched between two very different length configurations.
At first, the laser if feed back via the shorter loop, while the second
acts as an optical memory storing the output of the laser (configuration
A). Then, the shorter loop is excluded and the longer starts to feed
the laser, injecting the field stored before (configuration B). The
results show that the laser can exhibit different and independent
responses to the same drive depending on the dynamical regime, indicating
different levels of consistency; this can be see also by comparing
the spatiotemporal patterns in the configurations A-B. A transverse
Lyapunov analysis, carried out on the experimental data and compared
with simulations, leads to the conclusion that the consistent regimes
can be related to the presence of the weak chaos, and the inconsistent
ones corresponding to a strong chaos regime.

An experimental characterization of the mechanism for the emergence
of strong chaos was studied in \cite{Porte2014a}. The authors also
demonstrate certain similarity properties of the dynamics, relating
different pump currents regimes, to the feedback strength that can
be adjusted in such a way that the dynamics is similar.

\subsection{Reservoir computing}

Recently, time delayed systems with long time delays are used successfully
for reservoir computing \cite{Appeltant2011}. The space-time representation
for the processing response of the reservoir has been presented in
\cite{Larger2012,Martinenghi2012}, showing that it is indeed suited
to the matrix operation which is typically involved at the readout
layer of reservoir computing processing. The representation also well
illustrates how a delay-based reservoir computer emulates a network
of virtual neurons, 
\begin{figure}
\begin{centering}
\includegraphics[width=0.3\textwidth]{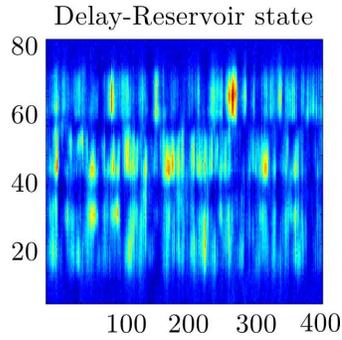}
\par\end{centering}

\caption{\label{fig:Larger2012} Space-time representation of reservoir computing
network input; the color encodes the amplitudes of the signals, with
red (blue) corresponding to large (small) values. The horizontal axis
is the delay time, the vertical the number of delays. (The figure
is kindly provided by L. Larger, similar to that in \cite{Larger2012})}
\end{figure}
with which the reservoir computing concept can be implemented in delay
systems. In Fig.\ref{fig:Larger2012}, a space-time plot is presented
for the processing of a spoken digit (''eight''), in a speech recognition
task.

\subsection{Chimera states \label{sub:Chimera-states}}

The increasing interest in the investigation of complex systems, either
spatially extended or on a network, has lead recently to the introduction
of previously unexpected states, named \textit{chimeras}. Such states
are characterized by the coexistence, in a large set of identical
oscillators, of two separated groups evolving synchronously and incoherently
respectively despite a homogeneous coupling \cite{Kuramoto2002}. 

\begin{figure}
\centering{}\includegraphics[width=0.75\columnwidth]{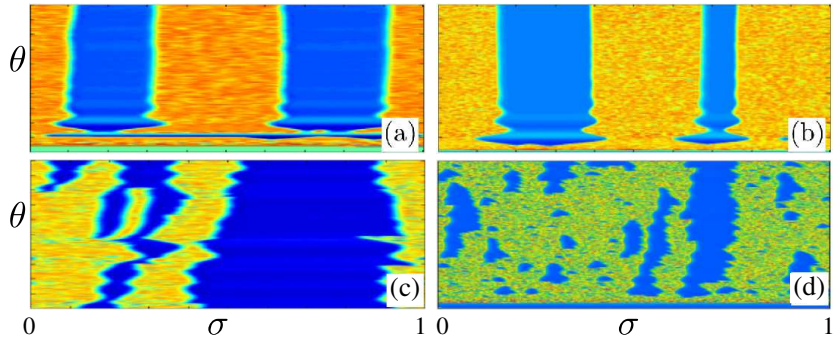}\caption{\label{fig:Larger2013} (from \cite{Larger2013} and \cite{Larger2015},
color online) Two-headed chimera, (a) experiment and (b) numerics.
Turbulent motion, (c) experiment and (d) numerics. (Reprinted (figure)
with permission from {[}L. Larger, B. Penkovsky, Yu. Maistrenko, Phys.
Rev. Lett. 111, 054103, 2013.{]} Copyright (2013) by the American
Physical Society.) }
\end{figure}

In \cite{Larger2013,Semenov2016}, time delayed systems are found
to display temporal patterns which split into regular and chaotic
components repeating within a delay unit; in the corresponding pseudo
space-time representation, the behavior is found to show the features
of a chimera state (see Fig.~\ref{fig:Larger2013}). This phenomenon
results from a strongly asymmetric nonlinear delayed feedback driving
a highly damped harmonic oscillator dynamics. The physical experiment
corresponds to a bandpass frequency modulation (FM) delay oscillator.
Numerous virtual chimera states are obtained and analyzed, through
experiment, theory, and simulations.

The report on a highly controllable experiment on chimeras, based
on an optoelectronic delayed feedback applied to a wavelength tunable
semiconductor laser, is presented in \cite{Larger2015}. There, a
wide variety of chimera patterns are investigated and interpreted.
A cascade of higher-order chimeras as a pattern transition from increasing
number of clusters of chaoticity is uncovered. Moreover, as the gain
increases, the chimera state is gradually destroyed on the way to
apparent turbulence-like system behavior.

\subsection{Optical cavity dynamics of localized excitations}

The study of optical cavity dynamics can be approached in two different
ways. The first is to use the Maxwell-Bloch equations with a suitable
modeling of the nonlinear processes involved in the gain medium and
imposing appropriate boundary conditions which amounts to provide
a set of non-local conditions (delay difference map) for the field.
The second way is to introduce a phenomenological model based on delay
equations which can represent an approximate description of the dynamics,
as multiple reflections at retarded times occur. A typical example
of the latter is the Lang-Kobayashi model, which takes into account
only the first of the multiple (in principle, infinite) reflections
from the distant mirror assuming a small feedback gain. 

Examples of complex dynamics in laser cavities are reported in the
literature. However, only recently the progresses in high temporal
resolution acquisition allowed to follow the microscopic evolution
of the field, possibly representing it in a spatio-temporal representation
and indicating the possibility of a description in terms of delay
models. 

\begin{figure}
\centering{}\includegraphics[width=0.75\columnwidth]{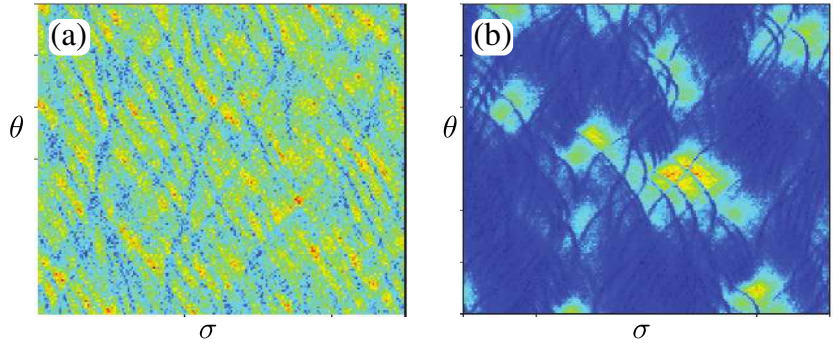} \caption{\label{fig:Turitsyna2013} (from \cite{Turitsyna2013}, color online)
Coherent structures in spatiotemporal dynamics in experiment for laminar
and turbulent regimes of fiber laser. Space-time diagram of intensity
for the laminar (a) and turbulent regime (b). Adapted by permission
from Macmillan Publishers Ltd: Nature Photonics (G.~Turitsyna et
al. ''The laminar-turbulent transition in a fibre laser'', Nat. Photon.
7, 10, pp. 783\textendash 786.), copyright (2013). }
\end{figure}

A fiber laser operating in laminar and turbulent regimes is investigated
in \cite{Turitsyna2013}. The transition from a linearly stable coherent
laminar state to a highly disordered state of turbulence is shown
as spatial size or the strength of excitation increases. In the paper,
it is evidenced that the laminar phase is analogous to a one-dimensional
coherent condensate using a spatio-temporal representation (see Fig.~\ref{fig:Turitsyna2013})
and that the onset of turbulence is due to the loss of spatial coherence.

\begin{figure}
\centering{}\includegraphics[width=1\columnwidth]{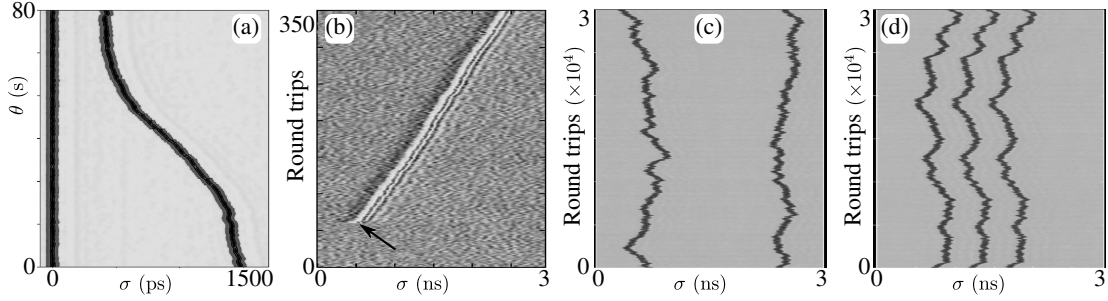} \caption{\label{fig:localized} (from \cite{Jang2013,Garbin2015,Marconi2015})
(a): Long-range interactions of a pair of cavity solitons. The temporal
evolution of two cavity solitons is shown in a spatio-temporal representation:
the horizontal axis is the cavity time variable, the vertical is the
number of cavity roundtrips. The initial separation is 1500 ps (577
soliton widths), the interaction is attractive until a stable separation
of 420 ps is attained \cite{Jang2013}. (b): Nucleation of phase bits
in space-time representation \cite{Garbin2015}. A phase perturbation
is applied (shown by arrow) while the system is in a stable stationary
locked state. Following that perturbation, a pulse is nucleated and
repeats with a periodicity close to the feedback delay time. (c) and
(d): Space\textendash time representation of the motion of dissipative
solitons (DS) in different situations \cite{Marconi2015}. (c): Two
independent DS. (d): A 3-DS molecule. Adapted by permission from Macmillan
Publishers Ltd: Nature Photonics (J.~K.~Jang et. al., ''Ultraweak
long-range interactions of solitons observed over astronomical distances'',
Nat. Photon. 7, 8, pp. 657\textendash 663.), copyright (2013); Nature
Communications (B.~Garbin et al., ''Topological solitons as addressable
phase bits in a driven laser'', Nat. Commun. 6, pp. 5915), copyright
(2015); Nature Photonics (M. Marconi et al., ''Vectorial dissipative
solitons in vertical-cavity surface-emitting lasers with delays'',
Nat. Photon. 9, pp. 450\textendash 455), copyright (2015).}
\end{figure}

A report about a weak interaction of solitons is presented in \cite{Jang2013}.
The experiment studies temporal optical cavity solitons injected and
recirculating in a coherently driven passive optical-fiber ring resonator.
Pairs of solitons are observed interacting over a range as large as
8,000 times their width. In the most extreme case, their temporal
separation changes as slowly as a fraction of an attosecond per roundtrip
of the 100-m-long resonator, or equivalently 1/10,000 of the wavelength
of the soliton carrier wave per characteristic dispersive length.
The interactions are so weak that, at the speed of light, an effective
propagation distance of the order of an astronomical unit can be required
to reveal the full dynamical evolution (see Fig.~\ref{fig:localized}(a)).
A model is discussed also, employing a modified Lugiato-Lefever equation
\cite{Lugiato1987} using the slow (number of roundtrips) and fast
(roundtrip variable) times as equivalent time and space variables
respectively.

It is shown both theoretically and experimentally in \cite{Marconi2014}
how lasing localized solitons form out of a passive mode-locked semiconductor
laser. The way the pulses become localized, and can be used for storing
information as independent addressable bits within a cavity trip time
is presented and discussed.

In \cite{Garbin2015} an experimental and analytical demonstration
is reported of the existence of longitudinal localized states in the
phase of laser light. They represent robust and versatile phase bits
that can be individually nucleated (see Fig.~\ref{fig:localized}(b))
and canceled in an injection-locked semiconductor laser, operated
in a neuron-like excitable regime and subjected to delayed feedback.
The observations are analyzed in terms of a generic model, which confirms
the topological nature of the phase bits and discloses their analogy
with Sine\textendash Gordon solitons.

The nonlinear polarization dynamics of a vertical-cavity surface-emitting
laser placed in an external cavity lead to the emission of temporal
dissipative solitons \cite{Marconi2015}. The vectorial character
is a consequence of the fact that they appear as localized pulses
in the polarized output, but leave the total intensity constant. When
the cavity roundtrip time is much longer than the soliton duration,
several independent solitons as well as bound states (molecules) may
be hosted in the cavity (see Fig.~\ref{fig:localized}). All these
solitons coexist together and with the background solution. The experimental
results are well described by a theoretical model that can be reduced
to a single delayed equation for the polarization orientation, which
allows the vectorial solitons to be interpreted as polarization kinks.
A Floquet analysis is used to confirm the mutual independence of the
observed solitons.

Recently, a photonic regenerative memory based upon a neuromorphic
oscillator with a delayed self-feedback has been shown and discussed
in \cite{Romeira2016}. The existence of a temporal response characteristic
of localized structures enables them as bits in an optical buffer
memory aimed to storage and reshaping of data information. The experimental
implementation, based upon a nanoscale nonlinear resonant tunneling
diode driving a laser, is supported by the analysis of the paradigm
of neuronal activity, the FitzHugh-Nagumo model with delayed feedback.

\section{Conclusions and perspectives}

\label{sec:concl}

The dynamics of systems with feedback is invariably modified by the
presence of a further timescale, related to the signal propagation
in the feedback loop. In many situations, the evolution is on very
slow scales and such effect is negligible; however, this is not always
the case. From optical setups, to neuroscience and complex networks
a relevant delay in the re-entering informations often causes sensible
effects such as multistability, quasi-periodicity, and several routes
to chaos. 

Particularly for long delays the observed behaviors appear high-dimensional.
In that case, roughly defined as the situation when the delay time
is much longer than other internal timescales, completely new phenomena
can be observed where purely temporal phenomena manifest complex behaviors
at different timescales. 

The first studies recognized the value of the auto-correlation function
for the preliminary analysis of the time series. Multiple recurrence
peaks evidenced the relevant timescales involved: accordingly, a re-organization
of the data provided a successful comparison of the dynamics with
that of an equivalent spatially extended system. The spatio-temporal
representation obtained in such a way is a fast and useful tool for
identifying the features hidden in the data. The method has proven
to be very successful and many applications of it have been presented,
disclosing the interpretation of quite complicated temporal patterns
in terms of well established spatio-temporal phenomena. As examples,
defects, space-time intermittency, fronts, coarsening and nucleation,
pinning and unpinning, chimeras, solitons creation, propagation, clustering
and annihilation, spirals, etc. have been demonstrated both in models
and experiments on long delay systems. The spatio-temporal representation
also illustrates the systematic presence of a drift in the information
propagation. This element, signature of the causality principle, has
been suitably formalized using the maximum comoving Lyapunov exponent.

The proper characterization of long delay system has been the subject
of a copious amount of papers. Beyond phenomenological evidences,
strong efforts to define their properties have been pursued in terms
of statistical and dynamical indicators. Starting with simple discrete
models, the main features of the Lyapunov spectrum have been studied
in details showing the peculiarity of the long delay systems. In particular,
the Lyapunov spectrum and its scaling with the delay time has permitted
to establish the latter quantity as the system size, leading to a
clear analogy with spatially extended systems. A classification of
the instabilities found in long delay systems and their connection
to the equivalent ones in spatially extended systems has been discussed.
Besides, the scaling of the main nonlinear indicators such as Kolmogorov-Sinai
entropy and the Kaplan-Yorke dimension have also been computed.

In the study of Lyapunov spectrum, already in the earlier works and
in the recent years a very important topic appeared. The elements
of the discrete subset of the Lyapunov spectrum may behave differently
(the anomalous, strongly unstable spectrum) with respect to the rest,
featuring what is probably one of the most distinctive character of
the long delay systems: the strong and weak chaos regimes. This classification
scheme is at the very basics of the correspondence with a spatio-temporal
dynamics. Given the importance of the distinction between the strong
and weak chaos, some key questions remain open. Among them, can those
regimes be determined by different, and more direct criteria than
a Lyapunov analysis? We remark also, that in the event of the presence
of anomalous exponents the long delay systems can be considered as
the bridge between low dimensional and high-dimensional, spatio-temporal-like
systems. Further investigations in the features of such passage are
of importance for the subject. 

Another point to stress is the connection between the weak/strong
chaos and an associated synchronization transition in delay coupled
networks. Indeed, in the strong chaos regime the linearized model
is formally the same of that considered in the framework of chaos
synchronization to determine the sub-Lyapunov exponents. The possible
change of sign of these anomalous exponents could be therefore interpreted
as the signature of synchronization with the delay forcing, thus permitting
a meaningful description in terms of a smooth pseudo spatio-temporal
evolution.

The long delay systems are intrinsically multiscale: the existence
of the long feedback timescale is at the basis of the very features
of the dynamics, and represents the key factor for the whole picture.
A rigorous approach to the equivalence with spatially extended systems
is properly pursued with the aid of multiple timescales techniques.
The exact derivation of the Ginzburg-Landau normal form has been provided
in such a way, studyng several models with long delay in the vicinity
of a a destabilization. However, only in a limited number of cases
a rigorous proof has been given. A challenging task is to extend it
to other situations, with possible mappings to normal forms other
than the Ginzburg-Landau.

Recognizing the fundamental role of the multiple timescales in long
delays system leads naturally to a generalization of the approach
when more than a single feedback loop, and therefore one delay term,
is part of the system. It has been shown that the situation of diverse,
hierarchically-long delays can be treated in a very similar way. However,
the mapping with spatio-temporal systems becomes more complicated
and the space-like variables involved are related to the number of
delay times considered, each acting on well-separated scales. Even
in the simple case of two delays, very complicated temporal patternings
may occur; the suitable representation, now keeping into account a
vectorial drift, shows the occurrence of peculiar bidimensional phenomena
such as spirals and topological defects. How a general approach can
be pursued in different situations, and which are the possible phenomena
expected therein, is a topic of great interest given the growing effort
to describe system with complex connected interactions, e.g. biological,
social and financial networks. As a consequence, we would expect the
occurrence of the temporal analogue of even more complicated spatio-temporal
phenomena such as line-defects in 3D systems and their motions.

Delayed effects are widespread in many complex situations, and the
way the signal propagates could also be related to the instantaneous
configuration of the setup which can change with time. The field of
state-dependent delay systems represents a further, challenging topic
which could benefit as well of the present approach.

When real systems are investigated, the influence of noise is unavoidable,
and it is very important to study its effects on delay systems with
long delays. Many questions are still open, for example, how the generically
increasing multistability of delay systems can interplay with noise
and whether the observed peculiar scaling behaviors could be analogous
to those of noisy spatio-temporal systems.

A growing number of studies are appearing on long delay systems, many
of them are experimental investigations. The possibility to unveil
an elegant mechanism at the very basis of quite complicated temporal
sequences is a powerful tool and can greatly motivate the research.
We expect even more groups treating non-local interactions in a proper
way and considering the idea of a space-time equivalence. Many fields
could benefit from embracing this approach, paving the way for a systematic
classification of setups and physical phenomena into well-established
spatio-temporal, dynamical categories.

\paragraph*{Acknowledgments}

We acknowledge fruitful collaborations and discussions with many collegues;
among them, F. T. Arecchi, S. Barland, I. Fischer, M. Giudici, L.
Larger, S. Lepri, F. Marino, A. Mielke, A. Politi, G.P. Puccioni,
K. Schneider, M. Wolfrum, M. Zaks. We thank the German Research Foundation
(DFG) for financial support in the framework of the Collaborative
Research Center 910.

\bibliographystyle{elsarticle-num}

\end{document}